\documentclass[11pt,a4paper]{article}

\usepackage[a4paper,margin=1in]{geometry}
\usepackage{setspace}
\usepackage[utf8]{inputenc}
\usepackage[T1]{fontenc}
\usepackage{textcomp}
\usepackage{lmodern}
\usepackage{microtype}
\usepackage{easyReview}

\usepackage{amsmath,amssymb,amsfonts,amsthm,bm,mathrsfs,mathtools}

\usepackage{graphicx}
\graphicspath{{./}{figs/}}
\usepackage{subcaption}
\usepackage{booktabs}
\usepackage{multirow}
\usepackage{array}
\usepackage{xcolor}
\usepackage[dvipsnames]{xcolor}
\usepackage{caption}
\DeclareCaptionLabelSeparator{naturebar}{\enspace\textbar\enspace}

\usepackage[colorlinks=true,linkcolor=blue!50!black,
            citecolor=blue!50!black,urlcolor=blue!50!black]{hyperref}
\usepackage{cleveref}

\usepackage[numbers,sort&compress,super,square]{natbib}
\setcitestyle{numbers,super,square}

\newcommand{\charT}{ChaRT}
\DeclareRobustCommand{\figref}[1]{Fig.~\ref{#1}}

\title{\textbf{A Foundation Model for Large-Scale Wireless Network Planning , Operation and Optimization}}
\author{%
  Xinyu Qin$^{2,3}$, Wenqiang Pu$^{2,*}$, Hongcheng Dong$^{2,3}$, Bingsheng Peng$^{2,3}$, Ye Xue$^{4}$,\\Linzhi Shen$^{5}$, Qi Yan$^{5}$, Tsung-Hui Chang$^{1,2,3,*}$, Zhi-Quan Luo$^{1,2,3}$\\[2pt]
  \small $^{1}$ Shenzhen Loop Area Institute, Shenzhen, China\\
  \small $^{2}$ Shenzhen Research Institute of Big Data, Shenzhen, China\\
  \small $^{3}$ The Chinese University of Hong Kong, Shenzhen, China\\
  \small $^{4}$ Sun Yat-Sen University, Shenzhen, China\\
  \small $^{5}$ Huawei Technologies, Shenzhen, China\\
  \small $^{*}$Corresponding author. E-mail: \texttt{changtsunghui@cuhk.edu.cn, wpu@sribd.cn}\\
}
\date{}

\begin{document}
\maketitle
\thispagestyle{empty}

\begin{abstract}
Wireless cellular networks form the connective tissue of human society, sustained by a continuous physical dialogue between engineered infrastructure and its surroundings. Radio signals emitted from base stations traverse terrain, diffract around buildings and scatter through streets before reaching billions of users. Together, these interactions produce the city-wide radio environment on which every network decision rests. Shaping this environment through deployment and optimization determines the connectivity societies rely on, yet learning it effectively at city scale and generalizing across diverse cities and deployments remain open challenges. Here we answer positively by introducing \charT{}, a foundation model that learns transferable radio representations from measurement reports generated by deployed cellular networks. These reports provide abundant multi-cell, multi-beam observations without dedicated campaigns, forming a scalable data foundation for city-scale learning. \charT{} embeds beam-level angular structure, network hierarchy and propagation-regime diversity in its architecture, and is pretrained through context-aware masked beam modelling and self-distillation with channel-model-constrained augmentation. We pretrain \charT{} on over one billion reports comprising 18.2 billion beam-level observations from 3,503 cells in one city. With a single set of weights, \charT{} reconstructs radio environments in unseen cities and transfers to radio map construction, new-site prediction and network parameter tuning. With only 1\% of labelled data, it supports user localization, beam prediction, propagation scenario classification and estimation of the signal-to-interference-plus-noise ratio. The learned representation further enables beamspace clustering for reusable radio-grid construction. These results establish \charT{} as a transferable foundation for network-wide intelligence.
\end{abstract}

% =====================================================================
%  Introduction
% =====================================================================
\section*{Introduction}
Wireless cellular networks serve as the nervous system of modern society, connecting more than five billion subscribers and carrying communications, sensing and AI-driven services that underpin digital economies, autonomous transportation and industrial automation \cite{GSMA2026MobileEconomy}. Managing these networks across their full lifecycle, including initial deployment, daily operation, optimization and maintenance, is intrinsically challenging. Wireless-system behaviour is jointly shaped by physical propagation, engineering configuration and human activity, and evolves as these factors change. Conventional network-management practices, such as drive-test campaigns, trial-and-error parameter tuning and offline ray-tracing simulation, cannot scale to this coupled, city-wide complexity\cite{Li2022RealWorld}. The \emph{digital twin network} (DTN) has therefore emerged as a central paradigm for next-generation network management, using a virtual replica of the deployed radio environment to predict network behaviour and enable proactive optimization\cite{Khan2022,ITUT2022,Mihai2022DTSurvey,HuiDTN2023}. Realizing this vision requires a model that can learn how radio signals propagate through complex environments and transfer this capability between cities and deployments without being rebuilt for each setting. However, physics-based tools such as ray-tracing simulators and standardized radio-propagation models, commonly known as channel models, require detailed site descriptions or scenario assumptions\cite{Remcom2023,Sionna2025,QuaDRiGa2014,3GPP_38.901}, while task-specific neural models often remain tied to their training scenarios.
 
On the other hand, the success of foundation models in language\cite{Brown2020,Touvron2023}, vision\cite{Radford2021,Kirillov2023} and scientific domains spanning remote sensing\cite{SkySensePP2025}, cardiac monitoring\cite{CSFM2026} and genomics\cite{LucaOne2025} points to a potential path forward: pretrain a single large model on broad, diverse radio data, with the ability to adapt to downstream wireless network optimization tasks. Several recent \emph{wireless foundation models} (WFMs) have explored this direction at the physical layer, using channel state information (CSI), which characterizes how a transmitted signal is altered along a radio link, or raw I/Q samples, which record the amplitude and phase of the received waveform\cite{wifo2,AlikhateebLWM2024,CatakBERT4MIMO2025,SpectrumFM2025,ICWLM2025,LLM4WM2025}. These models are a valuable proof of concept, but they cannot underpin a network-level DTN for two structural reasons. On the \emph{data axis}, the core obstacle is dimensional scaling. CSI is a high-dimensional, time-varying quantity: a single-cell base station may carry hundreds of antenna elements, yielding a channel matrix of comparable dimension per user, and its values fluctuate rapidly with user movement and environmental change. Collecting real CSI across thousands of such cells is therefore prohibitively expensive in both time and effort. Existing WFMs are consequently trained almost entirely on simulated channels, inheriting the assumptions those simulators encode rather than learning from the deployed world\cite{wifo2,AlikhateebLWM2024,CatakBERT4MIMO2025}. On the \emph{model axis}, the core obstacle is representational scope. A single-link CSI measurement is by construction an observation of one transmitter-receiver pair in isolation, encoding the link's multipath structure produced by radio signals arriving along reflected, diffracted and scattered paths, but carrying no information about the surrounding network. Quantities that govern network performance require joint observations across multiple cells. This information is structurally absent from any single-link representation, regardless of model capacity.

Towards capturing network-level radio information, the 5G standard itself offers a natural starting point. As part of routine operation, every active user equipment periodically
reports the signal strengths observed from surrounding cells, a record standardized by 3GPP as a measurement report
(MR)\cite{3GPP_38.331}.
No additional hardware or expensive measurement campaign is required and billions
of MRs are generated daily as a natural byproduct of network operation  though positional information is not available.
Each MR sample contains joint observations across multiple cells and beams, providing spatial and directional cues that may support learning network-scale radio structure that
single-link CSI cannot provide. Though MR data capture rich multi-cell, multi-beam radio observations
across the entire network, whether and how to build a foundation model
that generalizes across cities and transfers to downstream network
management tasks is far from obvious.
Three challenges must be addressed: how to represent beam measurements
consistently across heterogeneous deployments, how to learn global
propagation structure from highly incomplete observations, and how to capture the multiscale dependencies that govern network-wide radio behaviour within a unified model.

\begin{figure*}[htbp]
\centering
\begin{subfigure}[t]{\textwidth}
\phantomcaption\label{fig:dataset_overview}
\raggedright\textbf{a}\par\vspace{2pt}
\includegraphics[width=\textwidth]{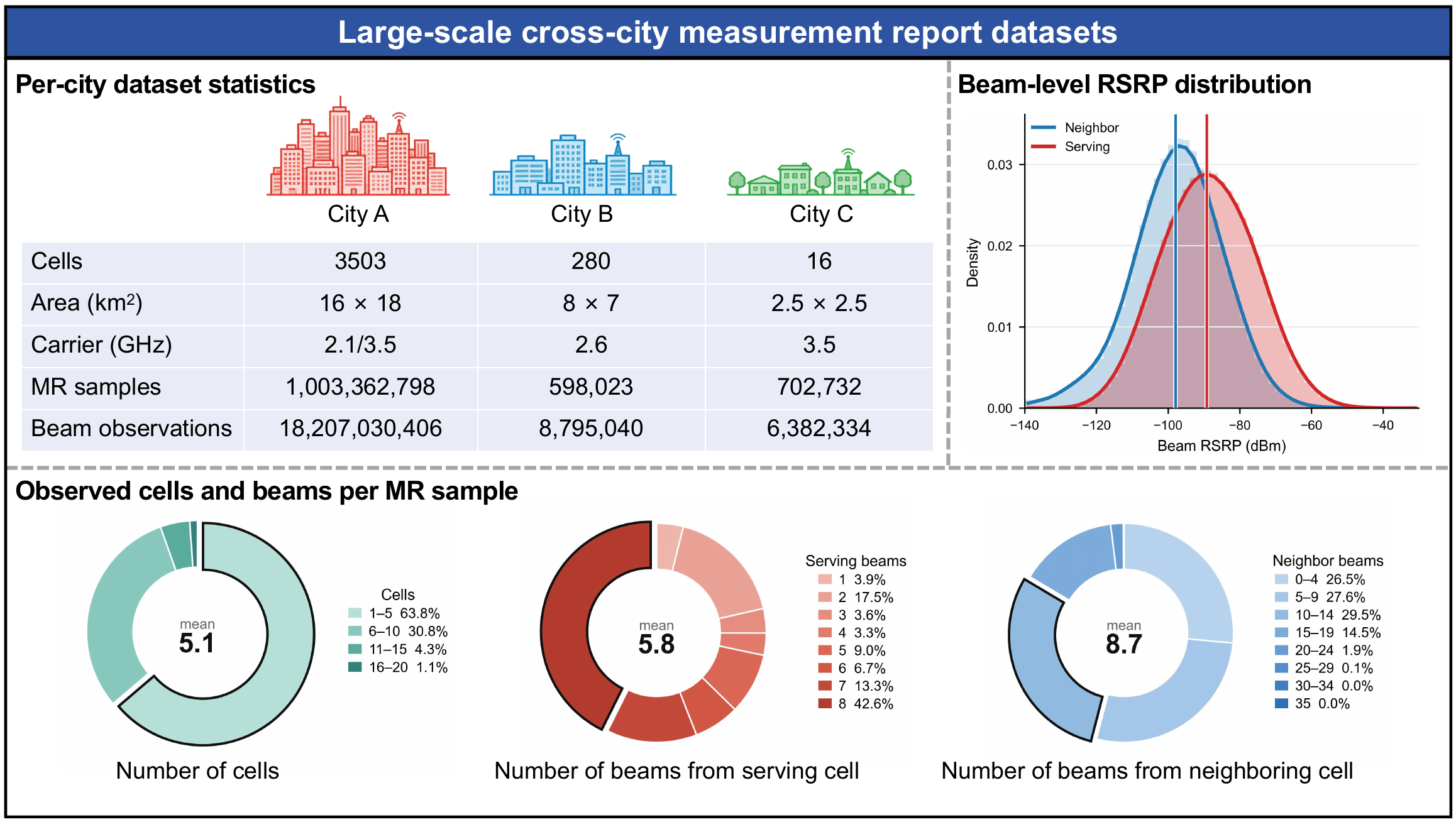}
\end{subfigure}

\vspace{0.8em}

\begin{subfigure}[t]{\textwidth}
\phantomcaption\label{fig:overview}
\raggedright\textbf{b}\par\vspace{2pt}
\includegraphics[width=\textwidth]{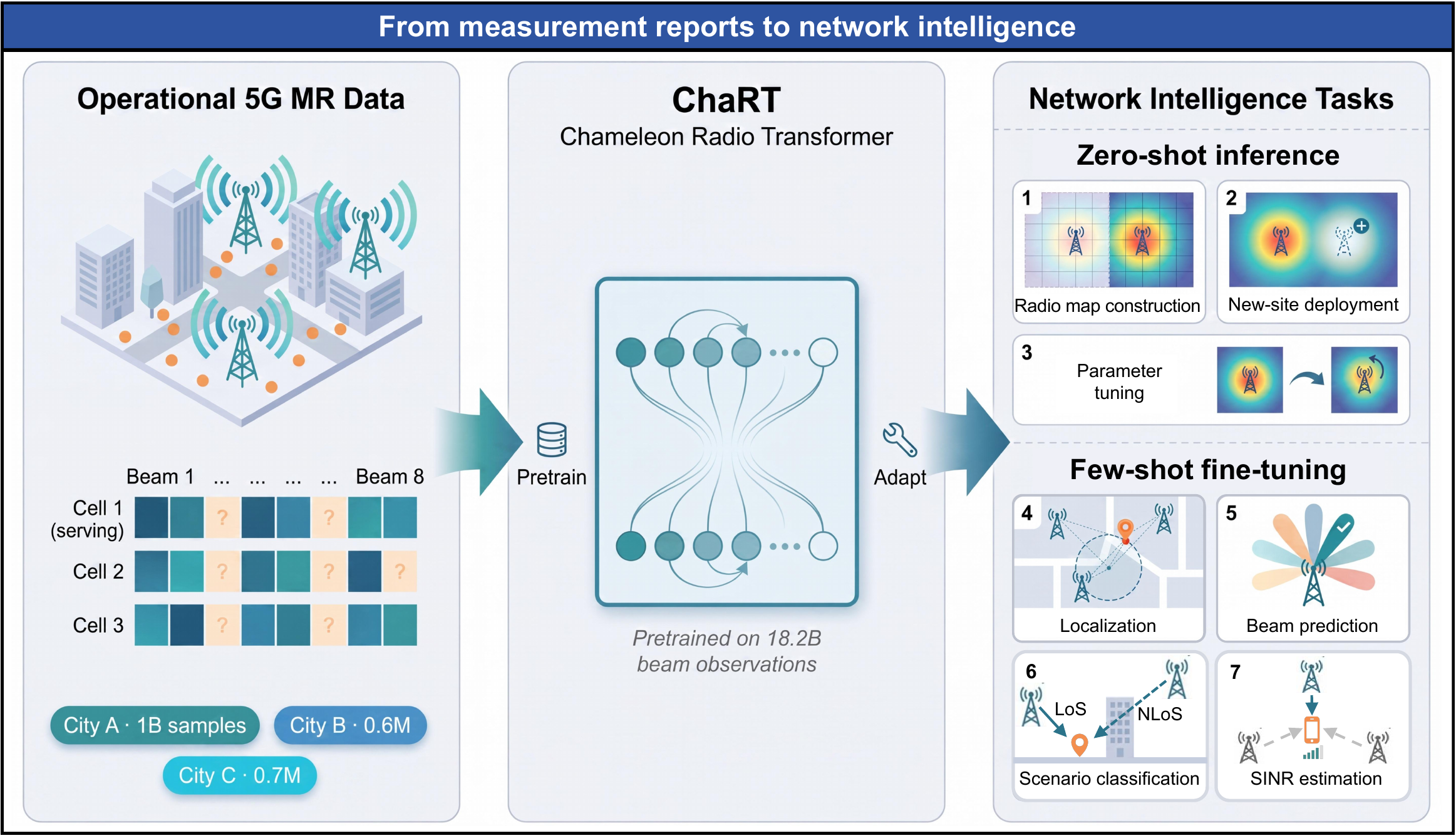}
\end{subfigure}
\caption{\textbf{Data foundation and overview of \charT{}.} \textbf{a,} Large-scale cross-city MR datasets and their statistical characteristics. Upper left, per-city dataset statistics.
Upper right, beam-level RSRP distributions for serving and
neighboring cells. Bottom, distributions of the numbers of
observed cells and beams per MR sample. \textbf{b,} \charT{} is pretrained and evaluated on 18.2 billion beam-level observations from deployed 5G networks in City A. Each MR sample records multi-cell, multi-beam RSRP observations that are tokenized and fed into the model. The frozen model directly serves zero-shot network planning, operation and optimization tasks (radio map construction, new-site deployment prediction, parameter tuning), or adapts to supervised downstream tasks (localization, beam prediction, scenario classification, SINR estimation) with 1\% of labelled data.}
\label{fig:fig1}
\end{figure*}

Here we present \textbf{\charT{}} (Chameleon Radio Transformer), to our knowledge the first wireless foundation model pretrained and evaluated on real-world, city-scale MR data, comprising over one billion samples and 18.2 billion beam-level observations collected from 5G networks in a single city (\figref{fig:dataset_overview}). Our results show that MR data contain transferable network-scale propagation structure: a single set of pretrained weights generalizes without retraining or architectural modification to cities, frequency bands and deployment scenarios not observed during pretraining (\figref{fig:overview}). At an 80\% masking ratio under cross-city zero-shot transfer, \charT{} reduces beam-level and cell-level RSRP reconstruction mean absolute error (MAE) by 25\% and 34\%, respectively, relative to XGBoost. In the cross-city new-site evaluation, \charT{} achieves $R^2=0.62$, compared with $R^2=0.24$ for TabPFN~v2 and negative values for XGBoost, Sionna and 3GPP TR~38.901. For antenna parameter tuning, it achieves $R^2$ values of 0.60--0.80 across five parameter types, compared with 0.37--0.48 for the strongest baseline. With only 1\% of locally labelled data, \charT{} reduces mean localization error from 111.0\,m to 62.9\,m, improves beam prediction accuracy from 66.9\% to 75.4\%, achieves a macro-F1 of 83.0\% for line-of-sight/non-line-of-sight (LoS/NLoS) classification and reduces the error of interference-aware link-quality estimation, measured by signal-to-interference-plus-noise ratio (SINR), from 2.82\,dB to 2.48\,dB. The same frozen representation also supports beamspace clustering to construct spatially compact and radio-homogeneous grids. These results provide evidence that the city-scale radio environment can be learned as a transferable substrate, opening a practical path towards a foundation model for network planning, operation and optimization using routinely collected network data.

\section*{Results}
A large-scale MR dataset from deployed 5G networks in China is collected to pretrain and evaluate \charT{}, spanning three major cities with distinct urban morphologies, network densities and frequency bands (\figref{fig:dataset_overview}). \textbf{City A} ($16{\times}18$\,km$^2$, 2.1/3.5\,GHz) is a dense urban macro deployment comprising 3,503 cells and over one billion MR samples. This single-city corpus is used for foundation model pretraining. \textbf{City B} ($8{\times}7$\,km$^2$, 2.6\,GHz) is a moderately dense urban deployment with 280 cells and 0.6 million samples at a distinct carrier frequency. \textbf{City C} ($2.5{\times}2.5$\,km$^2$, 3.5\,GHz) is a sparse suburban deployment with 16 cells and 0.7 million samples. Each MR sample observes a variable number of cells (mean 5.1) and beams---on average 5.8 beams from serving cells and 8.7 beams from neighboring cells per sample. The beam-level RSRP spans a wide dynamic range: beams from serving cells center around $-90$\,dBm, while beams from neighboring cells are distributed around $-98$\,dBm with a long tail below $-120$\,dBm (\figref{fig:dataset_overview}). In total, the dataset covers diverse urban environments, network scales and carrier frequencies, providing a practical testbed for evaluating the generalization of network-level wireless foundation models.

Built on this large-scale real-world dataset, we evaluate \charT{}
on the key capabilities that determine whether a foundation model can
serve the network management lifecycle. The ability to recover missing RSRP enables the construction of radio maps used in day-to-day network management. The ability to predict coverage for undeployed sites and antenna
reconfigurations enables planning and optimization decisions to be made before any physical change is executed. The ability to support tasks such as localization and beam management with minimal label data determines how efficiently the model can be deployed in a new operational environment.
Each of these currently requires dedicated measurement campaigns,
site-specific model retraining, or large amounts of label data,
making them expensive and difficult to scale across a live network.
We show that \charT{}, trained solely on real-world MR data, can
serve as a practical foundation for city-scale cellular network
management across cities and deployments entirely unseen during
pretraining.

% ------------------------------------------------------------------
%  Figures: beam-level and cell-level imputation
% ------------------------------------------------------------------
\begin{figure*}[htbp]
\centering
\begin{subfigure}[t]{\textwidth}
\phantomcaption\label{fig:beam_scenarios}
\raggedright\textbf{a}\par\vspace{2pt}
\includegraphics[width=\textwidth]{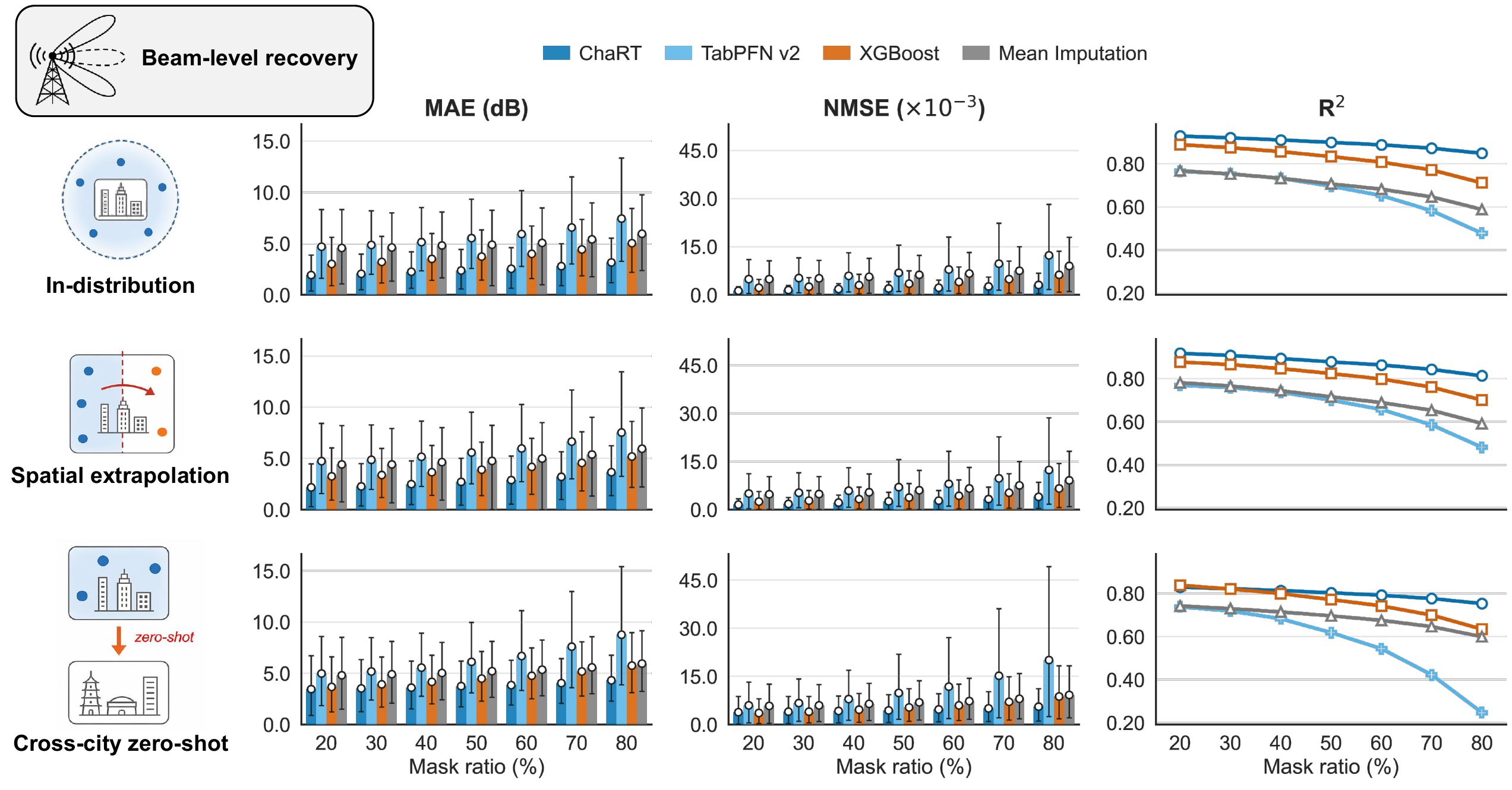}
\end{subfigure}
\par\vspace{0.6em}
\begin{subfigure}[t]{\textwidth}
\phantomcaption\label{fig:cell_scenarios}
\raggedright\textbf{b}\par\vspace{2pt}
\includegraphics[width=\textwidth]{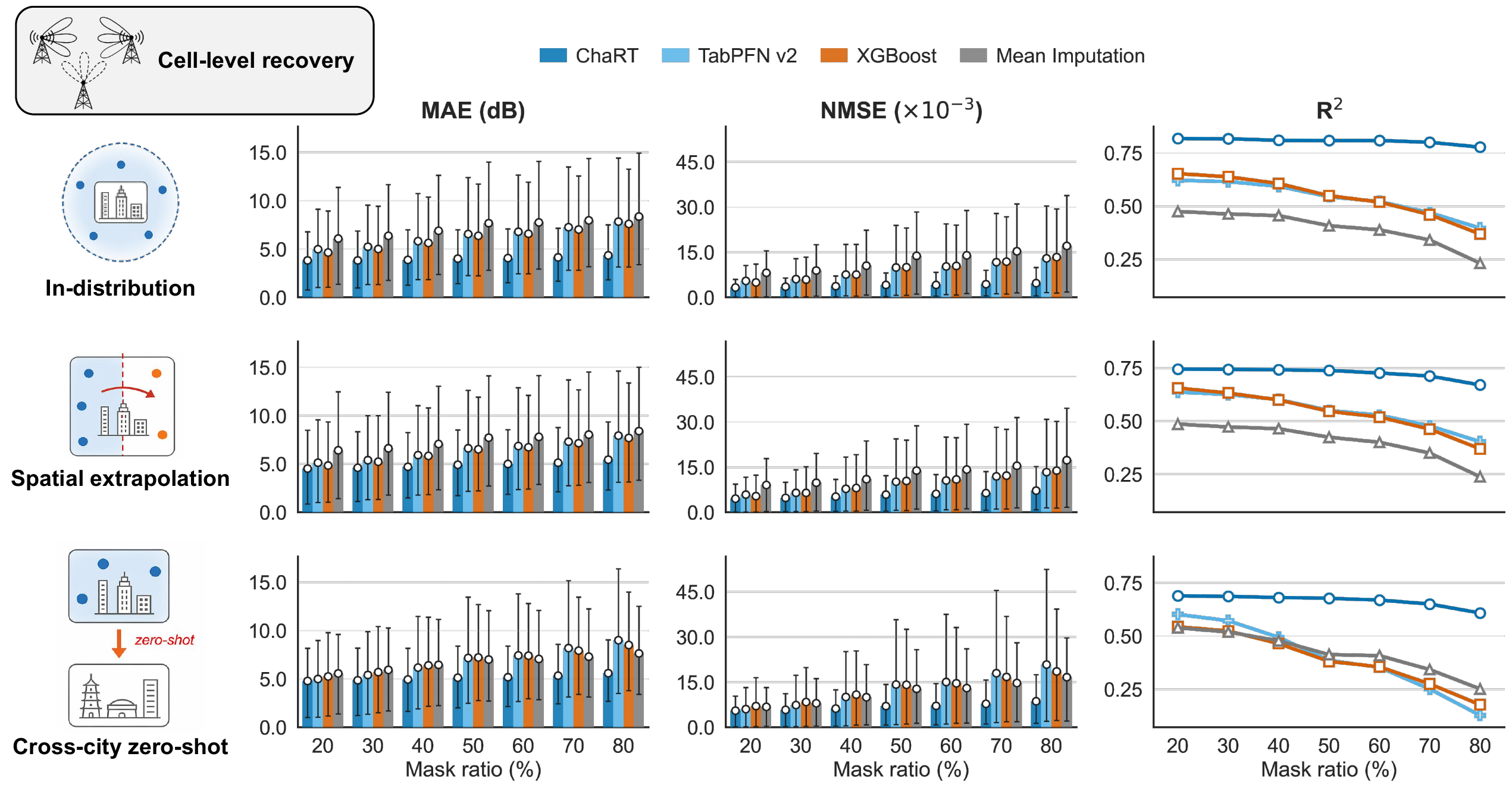}
\end{subfigure}
\caption{\textbf{Beam-level and cell-level RSRP reconstruction across three generalization settings.} \textbf{a,} Beam-level reconstruction performance as a function of the beam mask ratio. \textbf{b,} Cell-level reconstruction performance as a function of the cell mask ratio, where all beams of each selected target cell are removed. Lines show MAE, NMSE and $R^2$ for \charT{}, TabPFN~v2, XGBoost and mean imputation. Whiskers denote the 10th--90th percentiles of the per-sample error distribution. Rows correspond to in-distribution evaluation, spatial extrapolation and cross-city zero-shot transfer.}
\label{fig:reconstruction}
\end{figure*}

\subsection*{\charT{} reconstructs multi-beam RSRP from incomplete observations across cities}
Reconstructing multi-beam RSRP from incomplete observations is a prerequisite for radio map construction. We therefore evaluate whether \charT{} can reconstruct missing RSRP values from partial multi-cell, multi-beam context, which is the core foundation capability underlying the network lifecycle tasks considered below.

To disentangle the axes of generalization, we evaluate under three scenarios of increasing severity: in-distribution evaluation (random split within City A), spatial extrapolation (geographically disjoint City A regions) and cross-city zero-shot transfer (City B and City C, no fine-tuning). We compare \charT{} with mean imputation, XGBoost\cite{xgboost} and TabPFN v2\cite{tabpfn_v2}. Mean imputation fills a missing beam using the mean RSRP of the available beams in the same cell, or the available beams from the remaining cells when the target cell is fully missing; its performance indicates how much of the missing RSRP can be explained by the average signal level of the available context alone, without modeling beam-specific or inter-cell propagation structure. XGBoost is a strong gradient-boosted tree baseline, whereas TabPFN v2 is a transformer-based foundation model for tabular prediction. Both learning-based baselines receive the same tabular representation of observed RSRP values and physical configurations as \charT{}. We report MAE as the primary engineering metric, together with normalized mean squared error (NMSE) and coefficient of determination ($R^2$) as complementary measures of normalized reconstruction fidelity and explained variance, respectively. Whiskers denote the 10th--90th percentiles of per-sample errors and therefore characterize reconstruction reliability across heterogeneous radio conditions. We evaluate two complementary masking settings. In \emph{beam-level masking}, individual beams are hidden within otherwise observed cells, assessing recovery of fine-grained directional structure. In \emph{cell-level masking}, all beams from selected cells are removed simultaneously, requiring the model to infer an entirely unobserved cell from the remaining network context. Mask ratios range from 20\% to 80\%; higher ratios progressively suppress local evidence and place greater demands on global propagation reasoning.

\textbf{Beam-level recovery.}
Across all three scenarios, \charT{} achieves the lowest MAE at every masking ratio (\figref{fig:beam_scenarios}). Under in-distribution evaluation, MAE rises gradually from 1.93\,dB at 20\% masking to 3.17\,dB at 80\%, a 37\% reduction over XGBoost (5.05\,dB), 57\% over TabPFN v2 (7.44\,dB) and 47\% over mean imputation (5.95\,dB) at 80\% masking. The large gap relative to mean imputation confirms that the average signal level of the available context cannot account for directional variation among beams; accurate recovery requires beam-specific propagation structure.

Beyond average error, \charT{} substantially narrows the high-error tail.  At 80\% beam masking in the in-distribution setting, its per-sample MAE spans 1.20--5.52\,dB between the 10th and 90th percentiles, compared with 2.21--8.42\,dB for XGBoost, 3.27--13.33\,dB for TabPFN v2 and 2.38--9.77\,dB for mean imputation. The 90th-percentile error of \charT{} is 34\% lower than XGBoost and 59\% lower than TabPFN v2, indicating more reliable predictions in difficult propagation conditions rather than improvement only for well-observed samples.

The advantage persists under distribution shift. At 80\% beam masking, \charT{} achieves 3.65\,dB under spatial extrapolation and 4.32\,dB in cross-city transfer, corresponding to reductions of 30\% and 25\% relative to XGBoost (5.19\,dB and 5.75\,dB), and 51\% and 51\% relative to TabPFN v2 (7.52\,dB and 8.75\,dB). In cross-city transfer, \charT{} maintains $R^2 = 0.75$, compared with 0.63 for XGBoost, 0.25 for TabPFN v2 and 0.60 for mean imputation. At the mildest cross-city masking condition (20\%), XGBoost attains marginally higher $R^2$ and lower NMSE than \charT{} despite a higher MAE; this advantage disappears once masking exceeds 20\%. These results indicate that local correlations and sample-level signal averages suffice when dense context is available, whereas \charT{} becomes increasingly advantageous as the observable beam context diminishes---precisely the regime relevant to deployed cellular networks where beam observations are routinely sparse.

\textbf{Cell-level recovery.}
\charT{} provides its largest gains when entire cells are unobserved, a setting relevant to network planning and optimization where coverage must be predicted for new or reconfigured cells without direct measurements (\figref{fig:cell_scenarios}). At 80\% cell masking, \charT{} achieves MAEs of 4.35\,dB (in-distribution), 5.44\,dB (spatial extrapolation) and 5.61\,dB (cross-city), reducing error by 43\%, 29\% and 34\% relative to XGBoost and by 45\%, 31\% and 38\% relative to TabPFN v2.

\charT{} infers cell-specific coverage geometry, not merely a plausible average signal level. When every beam of a target cell is hidden, mean imputation retains some information about the overall RSRP scale but cannot distinguish one unobserved cell from another: its cross-city $R^2$ drops from 0.54 at 20\% masking to 0.25 at 80\%. \charT{} retains $R^2 = 0.61$ at the same ratio---a 144\% relative improvement---indicating that the learned representation captures cell-specific RSRP structure beyond the average signal level available in the observed context. The tail reliability advantage is equally clear: at 80\% cell masking in the cross-city setting, the 90th-percentile MAE of \charT{} is 9.03\,dB, versus 13.97\,dB for XGBoost (35\% reduction) and 16.39\,dB for TabPFN v2 (45\% reduction), confirming that \charT{} suppresses the large errors in weak-context or cell-edge conditions.

The beam- and cell-level results establish two complementary foundation capabilities: reconstructing directional beam structure within partially observed cells and inferring complete RSRP profiles for entirely unobserved cells. The latter is the basis for the zero-shot network planning and optimization tasks evaluated next.

\begin{figure*}[htbp]
\centering
\includegraphics[width=\textwidth]{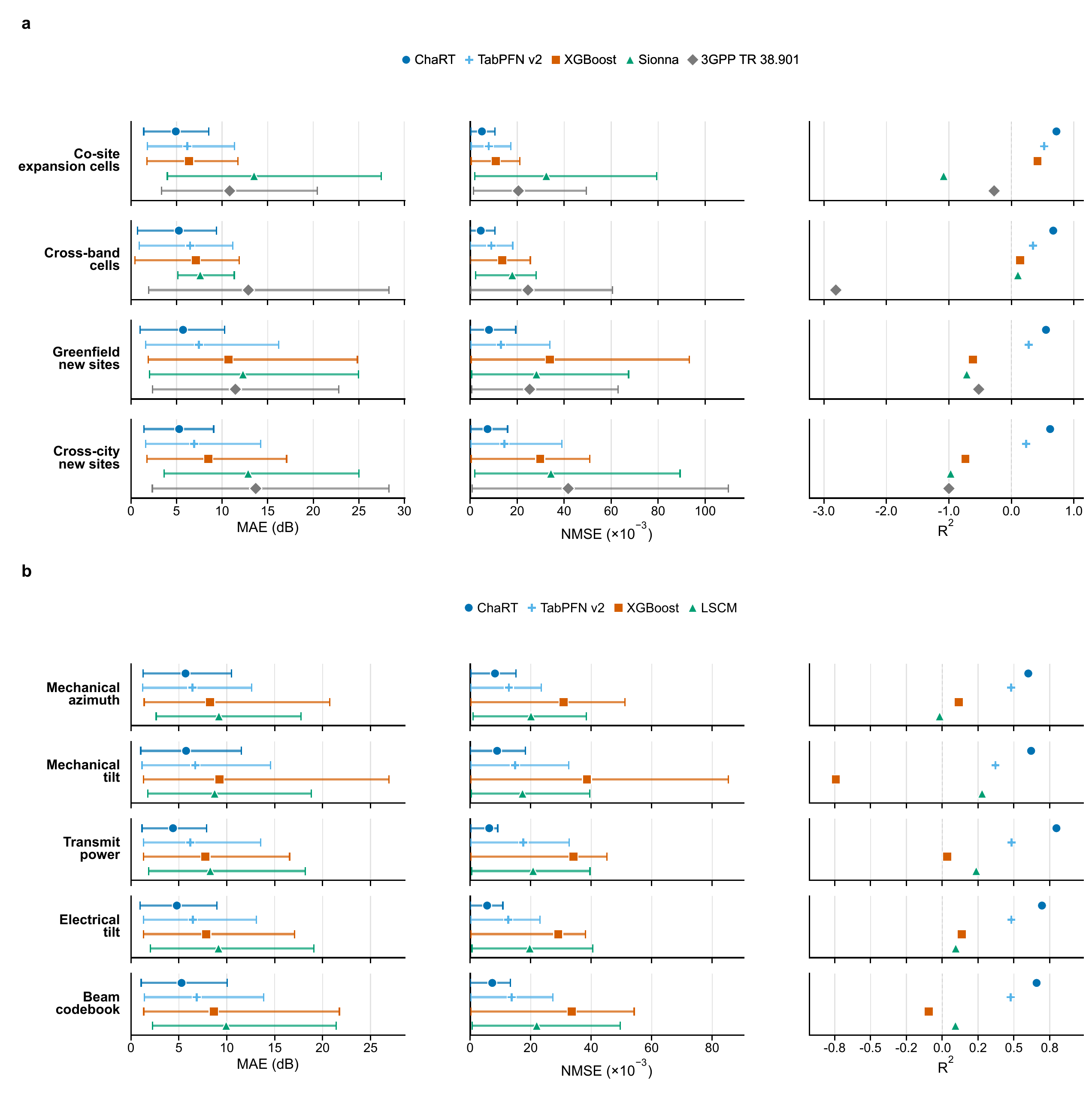}
\caption{\textbf{Zero-shot network planning and optimization.} \textbf{a,} New-site prediction under spatial extrapolation and cross-city transfer. The first three rows report co-site expansion cells, cross-band cells and greenfield new sites in City~A; the fourth row reports newly deployed base stations in City~B, which was not observed during pretraining. The comparison includes \charT{}, TabPFN~v2, XGBoost, Sionna and 3GPP TR~38.901. \textbf{b,} Prediction of post-tuning coverage in City~A after changes to mechanical azimuth, mechanical tilt, transmit power, electrical tilt and beam codebook. The comparison includes \charT{}, TabPFN~v2, XGBoost and LSCM. Points denote mean values and horizontal bars denote the 10th--90th percentiles where available.}
\label{fig:planning_optimization}
\end{figure*}

\subsection*{\charT{} transfers zero-shot to network planning and optimization}

Network planning and optimization require candidate actions, such as deploying a new cell, adjusting antenna parameters and reconfiguring beam codebooks, to be evaluated before physical implementation. Such counterfactual evaluation conventionally requires a dedicated measurement campaign or site-specific ray-tracing simulation for each change, neither of which scales to the hundreds of configuration changes made daily in a modern network. \charT{} addresses both tasks through the same cell-level masking mechanism: the target cell is masked and its full RSRP profile is predicted from the surrounding network context alone, without architectural modification and without labeled data. We evaluate two complementary tasks: new-site deployment prediction (planning) and antenna parameter tuning (optimization).

For new-site prediction, we evaluate under both spatial extrapolation and cross-city zero-shot transfer. In the spatial extrapolation setting, 210 cells from a geographically disjoint region of City A---unseen during pretraining---are masked at cell level. In the cross-city setting, 11 newly deployed base stations comprising 22 cells in City B are masked entirely. We compare against TabPFN~v2, XGBoost, the Sionna ray-tracing engine\cite{Sionna2025} and the 3GPP TR~38.901 statistical channel model\cite{3GPP_38.901}; all methods receive the same antenna configurations and network context.
For parameter tuning, we isolate the ability to generalize across antenna configurations using data from City~A collected over three temporal rounds, during which selected cells underwent parameter adjustments. The model is pretrained on Round~1 data only; test evaluation uses Round~2 and Round~3 data, where cells whose antenna parameters changed relative to Round~1 are masked at cell level and their post-tuning coverage is predicted from the surrounding context and updated configuration vector. We compare against TabPFN~v2, XGBoost and the localized statistical channel model (LSCM)\cite{lscm}, a physics-based data-driven approach that estimates the channel angular power spectrum from RSRP via sparse recovery and predicts post-tuning coverage through the modified beamforming pattern matrix. The anonymized UE coordinates used to render the maps are obtained from a location-tagged subset of MR samples collected through a consenting-user crowdsourcing programme\cite{Qin2026MRLSCM}; the corresponding measurements and predictions are spatially interpolated only for visualization.

\textbf{New-site deployment prediction.} \charT{} predicts the coverage of cells for which no target-site measurements are available, using only the surrounding network context. Under spatial extrapolation in City~A (\figref{fig:planning_optimization}\textbf{a}), performance is highest for co-site expansion cells ($R^2=0.72$, MAE\,=\,4.91\,dB), for which the target cell shares an existing base-station site with active cells. This is followed by cross-band cells ($R^2=0.67$, MAE\,=\,5.25\,dB), which operate at a different carrier frequency on an existing site, and greenfield new sites ($R^2=0.55$, MAE\,=\,5.71\,dB), where a new base station and its cells are introduced. This ordering is consistent with the amount of inter-cell context available for each deployment type.

In the greenfield setting, \charT{} achieves the highest coverage-prediction performance among the evaluated methods, with $R^2=0.55$ compared with 0.28 for TabPFN~v2. XGBoost, Sionna and 3GPP TR~38.901 yield $R^2$ values of $-0.62$, $-0.71$ and $-0.52$, respectively. Ray tracing depends on detailed three-dimensional building geometry, material properties and site-specific calibration, whereas the statistical channel model lacks the local environmental information required for directional coverage prediction. The performance of \charT{} is consistent with its pretrained representation capturing transferable relationships among antenna configuration, inter-cell geometry and received signal strength.

The advantage extends to cross-city transfer. For new base-station sites in City~B, which was not observed during pretraining (\figref{fig:planning_optimization}\textbf{a}), \charT{} achieves $R^2=0.62$ and an MAE of 5.28\,dB, corresponding to a 24\% lower MAE than TabPFN~v2 ($R^2=0.24$, MAE\,=\,6.92\,dB). XGBoost, Sionna and 3GPP TR~38.901 produce negative $R^2$ values. The radio maps in Extended Data Fig.~\ref{fig:radiomap_new_bs} show that the predictions of \charT{} more closely follow the observed beam directionality, distance-dependent signal decay and local spatial variation. These results indicate that \charT{} could support preliminary candidate-site screening while reducing reliance on measurement campaigns and per-site model calibration.

\textbf{Parameter tuning prediction.} \charT{} accurately predicts how coverage changes after antenna adjustment, enabling candidate configurations to be evaluated before deployment. We evaluate five parameter types from a real optimization campaign in City~A---antenna mechanical azimuth, mechanical tilt, transmit power, antenna electrical tilt and beam codebook---with the number of adjusted cells in each collection round summarized in Extended Data Table~\ref{tab:ed2} (\figref{fig:planning_optimization}\textbf{b}). Given only the post-tuning antenna configuration and surrounding network context, \charT{} achieves $R^2$ of 0.60--0.80 across all five types, with a 25--40\% relative gap in explained variance over the strongest baseline (TabPFN v2, $R^2$ of 0.37--0.48). Transmit power yields the highest accuracy ($R^2=0.80$, MAE\,=\,4.38\,dB), consistent with the expectation that a power change scales RSRP approximately uniformly. Antenna mechanical azimuth and mechanical tilt are more challenging because they redistribute radiated energy directionally; \charT{} retains $R^2$ of 0.60 and 0.62, whereas XGBoost yields negative $R^2$ and the LSCM remains below $R^2=0.28$. The LSCM establishes a statistical mapping between RSRP and the channel angular power spectrum, but recovering a high-dimensional sparse channel structure from low-dimensional and incomplete RSRP observations of a single cell is inherently ill-posed. \charT{} circumvents this limitation by conditioning on the multi-cell network context: the RSRP patterns of surrounding cells collectively constrain the propagation environment of the target cell, a relationship that the model has learned to exploit from millions of MR samples. The 10th--90th percentile MAE of \charT{} is roughly half that of XGBoost across parameter types, indicating more reliable predictions under diverse propagation conditions.

The radio map comparison in Extended Data Fig.~\ref{fig:radiomap_tuning} shows that \charT{} captures parameter-specific coverage effects. Each parameter type induces a distinct spatial signature: azimuth and tilt alter the main-beam orientation, transmit power changes the coverage footprint, and beam-codebook reconfiguration modifies the directional pattern. \charT{} closely reproduces both the direction and magnitude of these changes. The LSCM captures the directional shift of the beam pattern after reconfiguration but produces inaccurate absolute RSRP values because single-cell APS recovery is ill-posed. XGBoost generates noisy patterns, particularly around the building clusters visible in the overlaid footprints, whereas TabPFN v2 produces smooth but directionally inaccurate outputs. The inset magnifications further show that \charT{} tracks the local spatial redistribution of coverage after tuning, consistent with the model capturing parameter-dependent changes in the observed RSRP patterns.

Together, the results of the new-site and parameter tuning demonstrate that \charT{} has learned transferable radio propagation knowledge from real-world network data that generalizes across cities and antenna parameters. The evaluated baselines---representing the strongest available tabular, tree-based and physics-informed approaches--- all fail in settings that require joint reasoning over network context and antenna configuration. In practice, the coverage impact of a planned deployment or antenna configuration change can be estimated in a single forward pass, supporting candidate screening and reducing the measurement and calibration burden of conventional workflows.

\begin{figure*}[htbp]
\centering
\includegraphics[width=\textwidth]{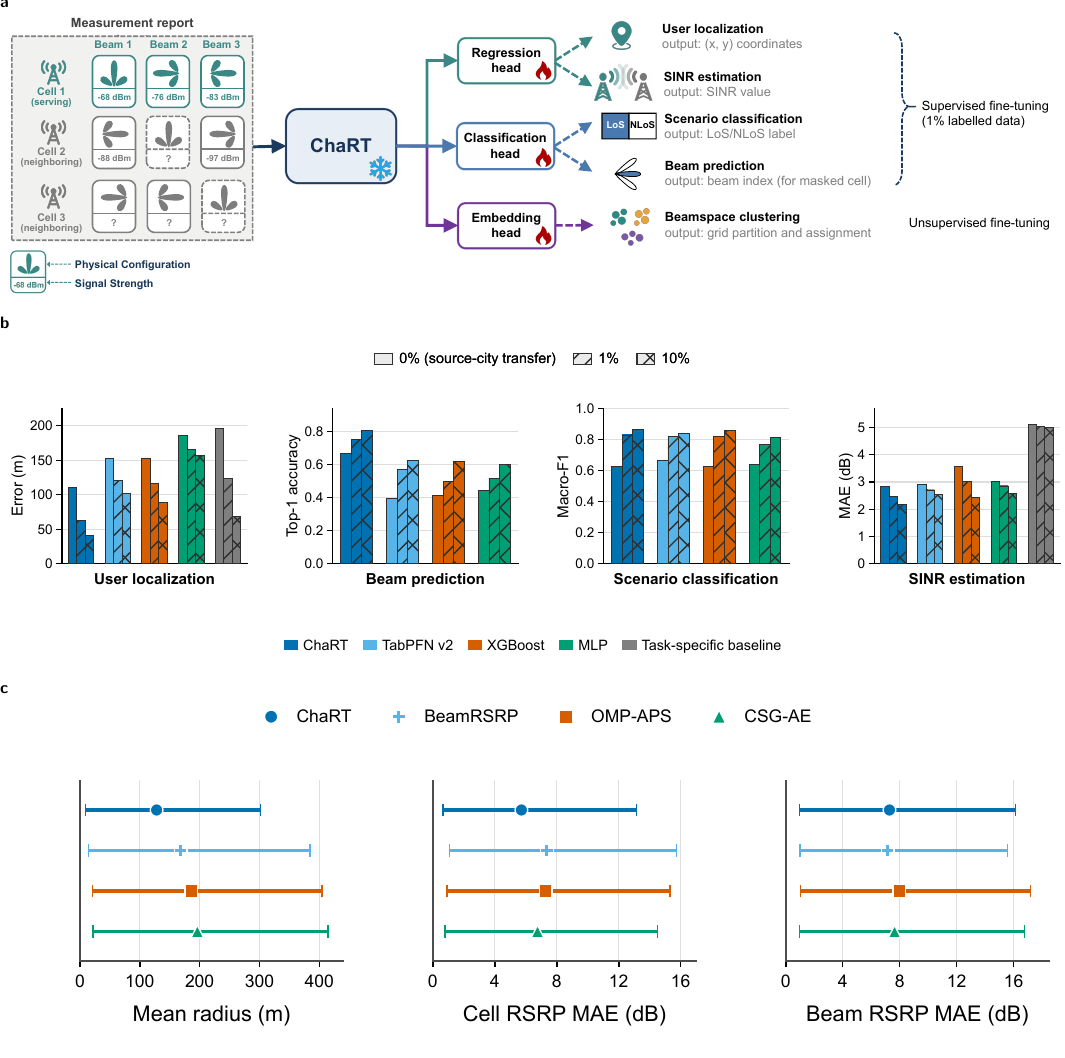}
\caption{\textbf{Downstream adaptation and evaluation of the frozen \charT{} representation.} \textbf{a,} The frozen backbone encodes MR observations into a shared representation. Lightweight regression, classification and embedding heads support four prediction tasks and beamspace clustering. The prediction heads are trained using labelled samples from City~A and then transferred to the target city, whereas the embedding head organizes MR samples into reusable radio grids. \textbf{b,} Cross-city performance for user localization, strongest-beam prediction, LoS/NLoS scenario classification and SINR estimation, shown from left to right. The task-specific baseline denotes WKNN for user localization and the RSRP-based parametric model for SINR estimation. \textbf{c,} Beamspace-clustering performance on held-out Round~3 data. Points denote means and horizontal bars denote the 10th--90th percentiles; lower values indicate better spatial compactness or radio homogeneity.}
\label{fig:downstream}
\end{figure*}

% ==================================================================
\subsection*{\charT{} empowers data-efficient adaptation to downstream tasks}

The pretrained representation supports five downstream applications through lightweight task-specific heads attached to the frozen backbone: user localization, beam prediction, propagation scenario classification, SINR estimation and beamspace clustering (\figref{fig:downstream}\textbf{a}). These heads are not included in backbone pretraining. For each of the four prediction tasks, the corresponding head is trained using labelled samples from City~A while the pretrained backbone remains frozen. Cross-city transfer is then evaluated by applying the City~A-trained head directly to the target city without target-city adaptation (0\%) or by further updating only the head using 1\% or 10\% of the labelled target-city samples. The 1\% setting is highlighted as the primary few-shot configuration. For beamspace clustering, an embedding head partitions the radio-data space into reusable radio grids for downstream network modelling and analysis.

For the four prediction tasks, we compare against three universal baselines---TabPFN~v2, XGBoost and a task-specific MLP---using identical input features. For localization, we additionally include a weighted $k$-nearest-neighbour (WKNN) fingerprinting method\cite{hu2024localization}; for SINR estimation, we include an RSRP-based parametric model\footnote{The parametric model estimates $\widehat{\mathrm{SINR}} = {\alpha_s \cdot 10^{\mathrm{RSRP}_s/10}} / ({\alpha_I\sum_{c\in\mathcal{N}}10^{\mathrm{RSRP}_c/10}+\sigma^2})$, where $\mathrm{RSRP}_s$ is the serving-cell power, $\mathcal{N}$ is the reported neighbour set, and $(\alpha_s,\alpha_I,\sigma^2)$ are fitted by least squares on the validation split.} that estimates SINR from the ratio of serving-cell power to aggregate neighboring-cell interference. The beamspace-clustering baselines and evaluation protocol are described together with the corresponding results below.

\textbf{User localization.} Accurate user localization from network-side measurements is essential for location-based services, yet collecting the dense fingerprint databases required by conventional methods is prohibitively expensive at city scale\cite{SalihuLocalization2024}. MR data do not include user locations by design because of protocol constraints and practical considerations, including privacy regulations. A small subset of MR samples can nevertheless be location-tagged through crowdsourcing programmes in which consenting users permit a proprietary application to periodically report GPS coordinates alongside signal measurements\cite{Qin2026MRLSCM}. These sparsely tagged samples provide the labels for localization. Without target-city adaptation, \charT{} achieves a mean localization error of 111.0\,m, which is 27\% lower than XGBoost (152.3\,m) and 41\% lower than MLP (186.6\,m). This result indicates that the pretrained representation already encodes coarse spatial structure from RSRP patterns alone (\figref{fig:downstream}\textbf{b}). With only 1\% of labelled data, the mean error decreases to 62.9\,m, a 46\% reduction relative to the strongest baseline at the same fraction (XGBoost, 116.8\,m). At 10\%, \charT{} reaches 40.9\,m, reducing the error by 40\% relative to WKNN (68.4\,m), 54\% relative to XGBoost (89.2\,m) and 60\% relative to TabPFN~v2 (101.6\,m). The pronounced improvement at 1\%, followed by diminishing returns at 10\%, indicates that pretraining already captures much of the relevant spatial structure and that labelled samples primarily calibrate the mapping from the learned representation to coordinates.

\textbf{Beam prediction.} Predicting the strongest beam for each user is central to beam management, initial access and handover in 5G networks\cite{AlrabeiahBeam2020}. Without target-city adaptation, \charT{} achieves a top-1 accuracy of 66.9\%, 22.6 percentage points above the strongest baseline (MLP, 44.3\%), indicating that radio-specific pretraining captures beam-level directional structure more effectively than the baselines can learn from the same input features (\figref{fig:downstream}\textbf{b}). With only 1\% of labelled data, \charT{} reaches 75.4\%, already exceeding all baselines trained with 10\% of the labelled data. At 10\%, \charT{} achieves 80.8\%, an 18.5-percentage-point advantage over the next-best method (TabPFN~v2, 62.3\%). The sustained advantage as more labels become available indicates that downstream supervision alone does not fully replace the directional structure acquired during pretraining.

\textbf{Scenario classification.} Distinguishing line-of-sight (LoS) from non-line-of-sight (NLoS) propagation is important for adaptive link management, handover optimization and channel-model selection, yet ground-truth propagation labels are rarely available in deployed cellular networks. Without target-city adaptation, all methods achieve comparable macro-F1 scores of approximately 62--66\% (\figref{fig:downstream}\textbf{b}). The distinction emerges in the data-scarce regime: with only 1\% of labelled data, \charT{} reaches 83.0\%, exceeding MLP (76.8\%) by 6.2 percentage points and TabPFN~v2 (82.0\%) by 1.0 percentage point. At 10\%, \charT{} reaches 86.7\%, compared with 86.1\% for XGBoost, 83.8\% for TabPFN~v2 and 81.5\% for MLP. The convergence of the learning-based methods at the larger data fraction suggests that LoS/NLoS propagation is separable given sufficient labels, whereas the advantage of \charT{} is concentrated in the regime where labelled samples are scarce.

\textbf{SINR estimation.} Accurate SINR estimation underpins link adaptation, scheduling and interference management, but requires joint information about the serving signal and interference from surrounding cells. The parametric SINR model, which estimates interference directly from linear-power RSRP ratios, yields an MAE above 5\,dB across all evaluated data fractions, indicating that raw power ratios alone are insufficient for accurate estimation (\figref{fig:downstream}\textbf{b}). 
Without target-city adaptation, \charT{} achieves an MAE of 2.82\,dB, compared with 2.91\,dB for TabPFN~v2, 3.03\,dB for MLP and 3.56\,dB for XGBoost. With 1\% of labelled data, the MAE decreases to 2.48\,dB---9\% below TabPFN~v2 (2.71\,dB) and 18\% below XGBoost (3.02\,dB). At 10\%, \charT{} reaches 2.18\,dB, maintaining a 10\% advantage over the strongest baseline (XGBoost, 2.42\,dB). Because SINR cannot be inferred reliably from single-cell observations alone, this result provides direct evidence that the pretrained representation captures network-level interference structure.

\textbf{Beamspace clustering.} Grouping MR samples with similar propagation conditions enables network optimization over radio grids rather than treating every measurement independently. We compare \charT{} with BeamRSRP, which clusters directly in beam-RSRP space; OMP-APS, which reconstructs the angular power spectrum from RSRP before clustering; and Channel Space Gridization Autoencoder (CSG-AE)~\cite{wang2026learning}. For this experiment, all methods construct $K=20{,}480$ grids. The \charT{} embedding head is refined using deep embedded clustering (DEC)~\cite{xie2016unsupervised} on Rounds~1 and~2, after which the learned grid centres are frozen for assignment and evaluation on Round~3. As shown in \figref{fig:downstream}\textbf{c}, \charT{} achieves the smallest sample-weighted mean within-grid radius of 128.2\,m, a reduction of 23.8\% relative to BeamRSRP. It also obtains the lowest CellRSRP MAE of 5.72\,dB, outperforming CSG-AE, OMP-APS and BeamRSRP by 1.04\,dB, 1.56\,dB and 1.64\,dB, respectively. BeamRSRP achieves the lowest BeamRSRP MAE because it clusters directly in the measurement space used for evaluation; nevertheless, \charT{} trails it by only 0.12\,dB while outperforming CSG-AE and OMP-APS by 0.35\,dB and 0.68\,dB, respectively. The t-SNE visualization in Extended Data Fig.~\ref{fig:cluster_tsne} provides a qualitative view of the representation structure. The representations learned by \charT{} form more compact and clearly separated groups, whereas the baseline representations exhibit greater inter-grid mixing. Together, these results show that \charT{} preserves beam-level radio homogeneity while producing radio grids that are more compact in physical space and more consistent in cell-level signal strength.

Across the four prediction tasks, \charT{} transfers from the same frozen backbone to continuous coordinates, binary propagation labels, discrete beam indices and scalar interference estimates, achieving consistent gains across all evaluated data fractions. Its advantage is most pronounced in the data-scarce regime, with 1\% of labelled samples already providing effective task adaptation. Beamspace clustering provides complementary evidence that the same representation can organize MR samples into spatially compact and radio-homogeneous grids. Together, these results indicate that \charT{} encodes a general-purpose description of the radio environment that can be efficiently specialized to diverse operational applications, rather than a task-specific feature space that must be reconstructed for each task.

\begin{figure}[htbp]
\centering
\begin{subfigure}[t]{0.32\linewidth}
\phantomcaption\label{fig:scaling_model}
\raggedright\textbf{a}\par\vspace{2pt}
\includegraphics[width=\linewidth]{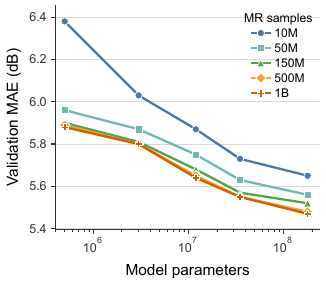}
\end{subfigure}
\hfill
\begin{subfigure}[t]{0.31\linewidth}
\phantomcaption\label{fig:scaling_data}
\raggedright\textbf{b}\par\vspace{2pt}
\includegraphics[width=\linewidth]{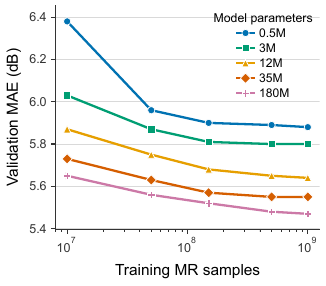}
\end{subfigure}
\hfill
\begin{subfigure}[t]{0.32\linewidth}
\phantomcaption\label{fig:scaling_grid}
\raggedright\textbf{c}\par\vspace{2pt}
\includegraphics[width=\linewidth]{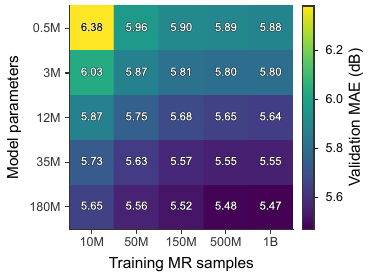}
\end{subfigure}
\caption{\textbf{Scaling of \charT{} with model size and pretraining data volume.} \textbf{a,} Validation MAE as a function of model size at five pretraining data scales ranging from 10 million to 1 billion MR samples. \textbf{b,} Validation MAE as a function of the number of pretraining samples for five model sizes. \textbf{c,} Joint scaling grid, with colour indicating MAE\,(dB). The MAE ranges from 6.38 to 5.47\,dB across the grid, corresponding to a total improvement of 0.91\,dB. Most of the improvement occurs in the small-model, low-data regime.}
\label{fig:scaling}
\end{figure}

\subsection*{\charT{} scaling approaches saturation with model size and same-city data volume}

\charT{} improves consistently as model size and pretraining data volume increase, although the marginal gains diminish in both cases (\figref{fig:scaling}). We vary model capacity from 0.5 million to 180 million parameters and pretraining data volume from 10 million to 1 billion MR samples drawn from the City~A corpus. The training and test regions are geographically disjoint, and generalization is evaluated in the held-out spatial region using a 40\% cell-masking ratio.

Increasing model capacity reduces MAE at every data scale, but the improvement per additional parameter decreases rapidly (\figref{fig:scaling}\textbf{a}). With 1 billion pretraining samples, increasing the model size from 0.5 million to 35 million parameters reduces MAE by 0.33\,dB, whereas a further increase to 180 million parameters yields an additional reduction of only 0.08\,dB. This trend is consistent across the data scales considered. Although the 180-million-parameter model outperforms the 35-million-parameter variant throughout the scaling grid, the average improvement is only 0.07\,dB.

Increasing the volume of same-city pretraining data produces a similar pattern of diminishing returns (\figref{fig:scaling}\textbf{b}). For the 35-million-parameter model, increasing the training set from 10 million to 150 million samples reduces MAE from 5.73 to 5.57\,dB, accounting for 89\% of the total improvement observed between 10 million and 1 billion samples. Performance changes little beyond 500 million samples, with an MAE of 5.55\,dB at both 500 million and 1 billion samples. The effect is also model-dependent: increasing the data volume by a factor of 100 reduces MAE by 0.50\,dB for the 0.5-million-parameter model but by only 0.18\,dB for the 35-million-parameter model. These results indicate that additional measurements from the same city provide progressively less information as the corpus grows. The joint scaling grid suggests limited further improvement beyond approximately 35 million parameters and 500 million MR samples under the evaluated setting (\figref{fig:scaling}\textbf{c}).

The observed saturation suggests that scaling efforts should increasingly prioritize data diversity once the benefits of larger models and additional same-city samples begin to diminish. Extending pretraining to additional cities, frequency bands and deployment configurations may expose the model to a broader range of propagation conditions than further increasing the sampling density within one city. Such expansion is operationally feasible because MRs are standardized by 3GPP and generated routinely in deployed cellular networks.

\section*{Discussion}

\charT{} advances wireless machine intelligence from individual-link and task-specific modelling towards transferable learning from network-scale observations in deployed cellular systems. By jointly encoding MRs and physical cell configurations, the model learns a shared representation that supports radio-environment reconstruction, network planning and data-efficient downstream inference. Its performance across the evaluated cities provides evidence that the coupled structure of cells, beams, configurations and propagation conditions can be learned within a common model rather than reconstructed separately for each site and application. More broadly, the results show how structured real-world network data can support transferable modelling of a large-scale wireless physical system when observations are paired with the configurations that generated them.

This shared representation provides a common predictive basis for network decision-making. Radio map construction, new-site assessment and parameter tuning can be expressed as different queries to the same model, while localization, beam prediction and interference-aware estimation require only lightweight task-specific heads and limited target-city supervision. In a future closed-loop system, operational measurements could update the model and network controllers could use its predictions to evaluate candidate actions. The present study establishes the predictive component of such a system, but integration with control policies and prospective evaluation of end-to-end network objectives remain open problems.

The present study also defines the scope of these conclusions. Pretraining uses data from City~A, while transfer is evaluated in two additional cities and across sub-6-GHz deployments, providing evidence across distinct network densities, frequencies and deployment conditions. Broader validation across regions and network architectures is still needed. MRs reflect the locations of active users and reporting policies rather than uniform sampling of the physical environment, which may under-represent low-traffic regions, weak-coverage areas and rare propagation conditions. In addition, \charT{} currently produces point predictions without calibrated uncertainty estimates. For parameter tuning, the retrospective evaluation demonstrates predictive accuracy under real-world network changes, while prospective interventions will be needed to quantify their causal effects. The current results support prediction, candidate screening and network decision support, with calibrated uncertainty and prospective validation forming the next steps towards autonomous closed-loop control.

Future work should extend pretraining across regions, frequency bands and deployment types to determine whether greater data diversity improves transfer under larger distribution shifts. Temporally indexed measurements, traffic states, building geometry and other environmental modalities could support dynamic modelling of how the radio environment evolves and responds to network interventions. Calibrated uncertainty, stronger physical constraints and out-of-distribution detection will be important for identifying when predictions can be trusted. Ultimately, prospective closed-loop trials are required to establish whether decisions informed by the learned representation improve network-level performance under real-world network changes.

% =====================================================================
%  Methods
% =====================================================================
\section*{Methods}

We design a unified methodology to pretrain a single model on large-scale 5G MR data such that it learns transferable representations of radio propagation across diverse network conditions. The methodology comprises five components: (i) a data pipeline that ingests over one billion standardized MR samples spanning 18.2 billion beam-level observations from deployed cellular networks in a city; (ii) a modeling framework that casts diverse network intelligence tasks---from radio map construction to network parameter tuning---as instances of a single masked beam prediction objective; (iii) a physics-informed model architecture that embeds beam-level angular structure, network hierarchy and propagation regime diversity into a unified tokenizer, attention mechanism and mixture-of-experts design; (iv) a contextual self-distillation pretraining strategy that unifies masked beam reconstruction and representation alignment within a student--teacher framework to prevent local interpolation collapse, and adopts channel-model-constrained augmentation via propagation invariants to further promote generalization across diverse network configurations; and (v) an application protocol through which the frozen model serves zero-shot operational tasks via different masking configurations, or adapts to supervised downstream tasks with minimal labeled data. The following sections detail each component. A progressive ablation under a
fixed spatial-extrapolation setting in City~A further isolates the contribution
of each architectural and pretraining component
(Extended Data Table~\ref{tab:ed3_ablation}).

\begin{figure*}[htbp]
\centering
\includegraphics[width=\textwidth]{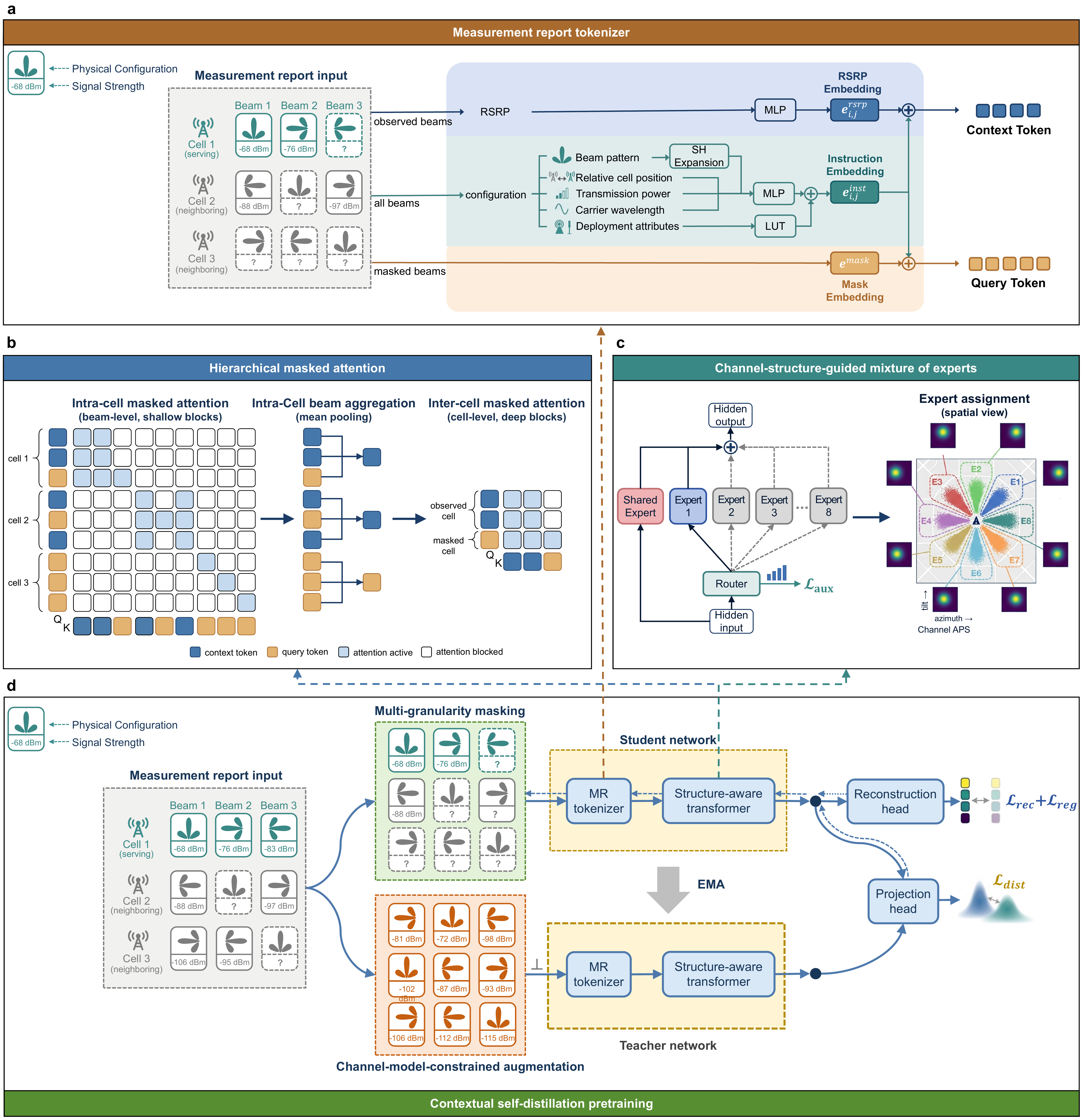}
\caption{\textbf{Architecture and pretraining framework of \charT{}.} \textbf{a,} The measurement report tokenizer encodes each beam's physical configuration and observed signal strength into a unified embedding, producing context tokens for observed beams and query tokens for masked or missing beams. \textbf{b,} Hierarchical masked attention captures intra-cell beam correlations in shallow blocks and inter-cell spatial dependencies in deep blocks, with masked entries serving only as queries. \textbf{c,} The channel-structure-guided mixture of experts routes each cell-level token to a single routed expert, while the shared expert processes all tokens. An auxiliary strongest-beam prediction loss, $\mathcal{L}_{\mathrm{aux}}$, supervises the router and aligns expert assignment with the dominant propagation direction. \textbf{d,} Contextual self-distillation pretraining combines multi-granularity masking for the student with channel-model-constrained augmentation for the teacher. The losses $\mathcal{L}_{\mathrm{rec}}$, $\mathcal{L}_{\mathrm{reg}}$ and $\mathcal{L}_{\mathrm{dist}}$ enforce beam reconstruction, missing-beam regularization and cross-view representation alignment, respectively.}
\label{fig:method}
\end{figure*}

%% ============================================================
\subsection*{Datasets}

Over one billion MR samples, comprising approximately 18.2 billion beam-level observations, were collected from deployed 5G networks in three cities in China. The dataset covers 3,799 cells operating at carrier frequencies of 2.1, 2.6 and 3.5\,GHz; per-city statistics are provided in \figref{fig:dataset_overview}. MRs are standardized by 3GPP\cite{3GPP_38.331} and are generated automatically by user equipment during routine network operation, including handover and periodic reporting, requiring no additional measurement campaigns or specialized infrastructure. Each sample records multi-beam RSRP values together with the identifiers of the serving and neighboring cells. The corresponding cell configurations, including antenna patterns, positions, transmission powers, carrier frequencies and deployment attributes (macro or non-macro, indoor or outdoor), are obtained offline and matched to the MR records. All UE identifiers are removed. Anonymized UE coordinates are retained only for the location-tagged crowdsourcing subset used for localization and map visualization; positional information is not available for the remaining MR samples.

%% ============================================================
\subsection*{Modelling}

\paragraph{MR observation model.}
Each MR sample is represented as an unordered set $$\mathcal{S} = \bigcup_{i=1}^{N}\{(\bm{c}_{i,j},\, y_{i,j})\}_{j=1}^{M_i}, $$ where $N$ is the number of detected cells, $M_i$ is the number of reported beams of cell $i$, $y_{i,j}$ is the RSRP value of the $j$-th beam, and $\bm{c}_{i,j}$ is a configuration vector encoding the physical identity of that beam---including the beam radiation pattern, relative cell position\footnote{The relative cell position is measured with respect to the serving cell in the same MR sample, reducing dependence on deployment-specific absolute coordinates.}, transmission power, carrier wavelength and deployment attributes. Several entries of $\bm{c}_{i,j}$ are shared across all beams of the same cell (relative position, power, wavelength, deployment attributes), while the beam radiation pattern differs per beam.

\paragraph{Channel-model basis.}
A key component of $\bm{c}_{i,j}$ is the beam radiation pattern $\bm{a}_{i,j}$, defined as the $j$-th row of the beamforming pattern matrix $\bm{A}_i$. Under the localized statistical channel model\cite{lscm}, the expected multi-beam RSRP vector {$\bm{y}_i \in \mathbb{R}^{M_i}$ relates to the channel angular power spectrum (APS) $\bm{x}_i \in \mathbb{R}^{N_A}$ via
\begin{equation}
\label{eq:lscm}
\bm{y}_i \;=\; \bm{A}_i\,\bm{x}_i,
\end{equation}
where $\bm{A}_i \in \mathbb{R}^{M_i \times N_A}$ is determined} by the antenna configuration and beam codebook, and $N_A$ is the number of angular bins. This linear relationship motivates two design choices in \charT{}: the tokenizer encodes each beam through its radiation pattern row $\bm{a}_{i,j}$, providing the model with explicit knowledge of each beam's angular selectivity (see ``Architecture''); and the pretraining augmentation recovers an approximate channel APS from observed beams via sparse recovery on \eqref{eq:lscm} \cite{lscm} to produce a physically consistent augmented view for the teacher network (see ``Pretraining \charT{}'').

\paragraph{Masked beam modelling objective.}
\charT{} is pretrained by a self-supervised objective termed \emph{context-aware masked beam modeling} (MBM), following the broader masked-reconstruction paradigm used in scalable representation learning\cite{He2022MAE}. A masking strategy partitions $\mathcal{S}$ into a visible context set $\mathcal{S}_c$ and a hidden target set $\mathcal{S}_t$. The model learns a predictor $f_\Theta$ whose architecture and pretraining procedure are described in the following sections. Given the observed subset $\mathcal{S}_c$ and the physical configurations of the target beams, the predictor estimates the masked entries. The parameters $\Theta$ are optimized by minimizing a composite objective over a training corpus $\mathcal{D}$:
\begin{equation}
\label{eq:mbm}
\min_\Theta\; \frac{1}{|\mathcal{D}|}\sum_{\mathcal{S}\in\mathcal{D}}\;\mathcal{L}\bigl(f_\Theta,\, \mathcal{S}_c,\, \mathcal{S}_t\bigr),
\end{equation}
where $\mathcal{L}$ combines reconstruction supervision with auxiliary regularization terms that enforce representation quality; the precise decomposition is given in the ``Pretraining \charT{}'' section. By pretraining on large-scale MR data through this unified objective, the model acquires a general understanding of radio propagation that can be directly applied or efficiently adapted to a wide range of downstream network intelligence tasks.

%% ============================================================
\subsection*{Architecture}

The architecture of \charT{} is shaped by three physical priors of cellular radio data (\figref{fig:method}): (i) each beam measurement probes a specific angular sector of the propagation environment through a known radiation pattern, motivating a \emph{physics-informed tokenizer} that encodes beam identity through its radiation pattern and cell geometry rather than a fixed cell or beam index (\figref{fig:method}\textbf{a}); (ii) radio propagation exhibits a natural two-level hierarchy---intra-cell beam correlations governed by the local scattering environment, and inter-cell dependencies shaped by spatial geometry---motivating a \emph{hierarchical masked attention} mechanism (\figref{fig:method}\textbf{b}); and (iii) cells across a network exhibit diverse channel structures that require specialized processing, motivating a \emph{channel-structure-guided mixture of experts} that routes cells to regime-specific sub-networks (\figref{fig:method}\textbf{c}).

\paragraph{Measurement report tokenizer.}
In MR data, the RSRP observed at each beam is jointly determined by the transmitting cell's physical configuration and the propagation environment. To enable the model to disentangle these factors and reason about how configuration changes affect coverage, the tokenizer explicitly binds \emph{what was observed}, namely signal strength, with \emph{which physical entity produced it}, namely the beam and cell identity. This design contrasts with standard sequence tokenizers that rely on positional encodings tied to a fixed ordering.

For each beam $j$ of cell $i$, two parallel embeddings are constructed (\figref{fig:method}\textbf{a}). For the RSRP branch, all measurements are converted to the logarithmic dBm scale before tokenization to compress their dynamic range and improve numerical stability. The \emph{RSRP embedding} then maps each scalar value $y_{i,j}$ to $\bm{e}^{\mathrm{rsrp}}_{i,j} \in \mathbb{R}^d$ through a multi-layer perceptron (MLP). The \emph{instruction embedding} $\bm{e}^{\mathrm{inst}}_{i,j} \in \mathbb{R}^d$ encodes the physical identity of the measurement. Discrete deployment attributes are mapped through learnable lookup tables (LUTs), while continuous attributes, including the beam radiation pattern $\bm{a}_{i,j}$, relative cell position $\bm{u}_i$, transmission power $P_i$ and carrier wavelength $\lambda_i$, are encoded through an MLP. Spherical harmonics (SH) provide a compact basis for representing complex antenna radiation patterns while retaining their angular structure\cite{Schmitz2012SphericalHarmonics}. The original pattern $\bm{a}_{i,j}$ is sampled on a $72\times91$ angular grid at intervals of $5^\circ$ in azimuth and $2^\circ$ in downtilt, yielding 6,552 values. Truncating its SH expansion at degree $L=15$ produces $(L+1)^2=256$ coefficients, reducing the dimensionality by a factor of 25 while preserving the angular information needed to distinguish beams.

For observed beams, the final embedding is $\bm{e}_{i,j}=\bm{e}^{\mathrm{inst}}_{i,j}+\bm{e}^{\mathrm{rsrp}}_{i,j}$, forming a context token. For masked or originally missing beams, the RSRP component is replaced by a learnable mask embedding $\bm{e}^{\mathrm{mask}}\in\mathbb{R}^d$, yielding $\bm{e}_{i,j}=\bm{e}^{\mathrm{inst}}_{i,j}+\bm{e}^{\mathrm{mask}}$ and forming a query token. Because each beam's identity is encoded through $\bm{c}_{i,j}$ rather than a fixed index, and the subsequent Transformer blocks operate without order-dependent positional encodings, the architecture is permutation equivariant\cite{Lee2019SetTransformer}. Reordering the cells or beams therefore reorders the corresponding outputs without changing their values, allowing \charT{} to accommodate the unordered and variable-length structure of MR samples.

\paragraph{Hierarchical masked attention.}
MR data exhibit a two-level physical hierarchy: within each cell, beam-level RSRP correlations are governed by the beamforming pattern and local multipath structure; across cells, dependencies reflect spatial propagation geometry and inter-cell interference. Learning these two levels of structure simultaneously is essential---beam-level recovery exploits intra-cell correlations to reconstruct individual missing beams, while cell-level prediction requires inter-cell reasoning to infer the radio environment of an entirely unobserved site. Rather than applying global attention across all beams---which conflates these two scales and incurs quadratic cost in total beam count---the encoder is decomposed into shallow \emph{intra-cell} blocks and deep \emph{inter-cell} blocks (\figref{fig:method}\textbf{b}).

In the intra-cell blocks, attention is restricted to beams within the same cell. A beam-level attention mask prevents masked beams from serving as keys while allowing them to attend to visible beams as queries, ensuring that each cell's representation is built exclusively from observed measurements without information leakage from masked entries. After the intra-cell blocks, per-beam hidden states are compressed into a single cell-level token via mean pooling: $\bm{h}_i = \frac{1}{M_i}\sum_{j=1}^{M_i}\bm{h}_{i,j}$. The deep inter-cell blocks then apply an analogous cell-level mask across cell tokens, enabling the model to reason about spatial relationships across the network. This two-level decomposition mirrors the physical hierarchy of cellular data and captures both intra-cell beam correlations and inter-cell propagation geometry within a single forward pass.

\paragraph{Channel-structure-guided mixture of experts.}
Cells across a network exhibit diverse channel structures, reflected in distinct dominant propagation directions and multipath profiles. A single shared feed-forward network must average over these heterogeneous regimes, limiting its capacity to capture structure-specific patterns. To address this limitation, \charT{} replaces the feed-forward network in each inter-cell Transformer block with a mixture-of-experts (MoE) module that routes each cell-level token according to its dominant channel structure (\figref{fig:method}\textbf{c}).

Inspired by the shared-expert isolation strategy of DeepSeekMoE~\cite{Dai2024DeepSeekMoE}, the MoE module comprises one shared expert $\mathrm{SE}(\cdot)$ and $E$ routed experts $\{\mathrm{RE}_{e}(\cdot)\}_{e=1}^{E}$, each implemented as a two-layer MLP. The shared expert processes every cell token to capture propagation patterns common across channel regimes, while a top-1 routing strategy activates only the routed expert with the highest gating probability to model structure-specific features:
\begin{equation}
\mathrm{MoE}(\bm{h}_i)=\mathrm{SE}(\bm{h}_i)+\mathrm{RE}_{e^{\star}}(\bm{h}_i),\qquad e^{\star}=\arg\max_{e\in\{1,\ldots,E\}}p_e(\bm{h}_i),
\label{eq:csg_moe}
\end{equation}
where $p_e(\bm{h}_i)$ is the routing probability assigned to expert $e$. A lightweight gating network projects each cell token into $E$ routing logits and applies a softmax operation to obtain the normalized probability distribution $p_e(\bm{h}_i)$ over the routed experts.

Standard MoE routing is learned solely from the end-task objective and therefore does not guarantee that expert specialization corresponds to physically meaningful propagation regimes. To align the routing decisions with channel structure, we supervise the gating network using an auxiliary strongest-beam prediction task. Because each beam illuminates a specific angular sector determined by its radiation pattern, the strongest-beam index provides a physical proxy for the dominant propagation direction. This auxiliary objective encourages each routed expert to specialize in cell-level tokens associated with coherent dominant propagation directions (\figref{fig:method}\textbf{c}, right). A spatial analysis of a representative cell provides qualitative evidence for this mechanism: compared with vanilla MoE without the auxiliary strongest-beam objective, CSG-MoE produces more contiguous, directionally organized expert regions that follow the strongest-beam partition (Extended Data Fig.~\ref{fig:ed_csg_moe}). It thereby partitions the representation space according to channel structure without requiring explicit environment labels. The corresponding auxiliary loss is defined in the pretraining objective below.

%% ============================================================
\subsection*{Pretraining \charT{}}

Standard masked reconstruction applied to sparse MR data risks degenerating into local interpolation, whereby the model learns to fill missing beams by averaging nearby observed values rather than encoding the global propagation structure needed for cross-city transfer. To overcome this limitation, \charT{} adopts a contextual self-distillation strategy that combines beam-level reconstruction with representation-level alignment across physically consistent views (\figref{fig:method}\textbf{d}).

\paragraph{Student--teacher framework.}
The predictor $f_\Theta$ introduced in \eqref{eq:mbm} is decomposed into three components: a backbone encoder $\mathcal{E}_\theta$ (the hierarchical transformer described above), a reconstruction head $h_\psi$ that maps encoder outputs to per-beam RSRP predictions, and a projection head $g_\phi$ that maps encoder outputs to a normalized embedding space for representation alignment. The full parameter set is $\Theta = \{\theta, \psi, \phi\}$.

During pretraining, two copies of the encoder are maintained: a \emph{student} $\mathcal{E}_{\theta_s}$ updated by gradient descent, and a \emph{teacher} $\mathcal{E}_{\theta_t}$ updated as an exponential moving average (EMA) of the student weights: $\theta_t \leftarrow \lambda\,\theta_t + (1-\lambda)\,\theta_s$, with decay $\lambda = 0.996$. Given an MR sample, the masking strategy produces a partial context $\mathcal{S}_c$ for the student; simultaneously, a channel-model-constrained augmentation produces a fully visible augmented view $\mathcal{S}'$ for the teacher. The student receives gradients from the composite loss; the teacher provides a slowly evolving, stable alignment target via its EMA-smoothed parameters.

\paragraph{Pretraining objective.}
The composite loss $\mathcal{L}$ in \eqref{eq:mbm} decomposes into four terms, each addressing a distinct learning requirement:
\begin{equation}
\label{eq:total_loss}
\mathcal{L}_{\mathrm{total}} \;=\;
\underbrace{\mathcal{L}_{\mathrm{rec}}}_{\text{beam reconstruction}}
\;+\; \beta\,\underbrace{\mathcal{L}_{\mathrm{dist}}}_{\text{representation alignment}}
\;+\; \alpha_1\,\underbrace{\mathcal{L}_{\mathrm{aux}}}_{\text{expert routing}}
\;+\; \alpha_2\,\underbrace{\mathcal{L}_{\mathrm{reg}}}_{\text{missing-beam penalty}},
\end{equation}
with $(\beta, \alpha_1, \alpha_2) = (1.0,\, 0.1,\, 0.1)$ throughout.

\emph{Masked beam reconstruction} ($\mathcal{L}_{\mathrm{rec}}$) drives the student to recover RSRP values of masked target beams from partial context:
\begin{equation}
\mathcal{L}_{\mathrm{rec}}\;=\!\!\!\sum_{(\bm{c},y)\in\mathcal{S}_t}\!\!\!
\bigl|\,y - h_\psi\!\bigl(\mathcal{E}_{\theta_s}(\mathcal{S}_c,\bm{c})\bigr)\bigr|.
\end{equation}

\emph{Representation alignment} ($\mathcal{L}_{\mathrm{dist}}$) prevents the model from relying on local interpolation by forcing the student's representation of the masked context to be globally consistent with the teacher's representation of the full augmented view. Both encoder outputs are projected into a normalized space by projection heads $g_{\phi_s}$ and $g_{\phi_t}$, converted to probability distributions via temperature-scaled softmax, and compared by cross-entropy:
\begin{equation}
\mathcal{L}_{\mathrm{dist}} = -\sum_{k=1}^{K} P_t^{(k)} \log P_s^{(k)},
\end{equation}
where $P_s$ and $P_t$ are student and teacher distributions over $K$ dimensions with temperatures $\tau_s$ and $\tau_t$ ($\tau_t < \tau_s$). The teacher distribution is centered by a batch-level EMA bias before the softmax to prevent representational collapse\cite{Caron2021}. Together, $\mathcal{L}_{\mathrm{rec}}$ provides fine-grained RSRP-level supervision while $\mathcal{L}_{\mathrm{dist}}$ enforces global semantic consistency, encouraging the model to encode propagation structure rather than local interpolation shortcuts.

\emph{Expert routing} ($\mathcal{L}_{\mathrm{aux}}$) is a cross-entropy loss that supervises the CSG-MoE gating network via the strongest-beam prediction task described above:
\begin{equation}
\mathcal{L}_{\mathrm{aux}} = -\sum_{e=1}^{E} \mathbb{I}(b^* = e)\,\log\,p_e(\bm{h}_i),
\end{equation}
where $b^*$ is the index of the strongest observed beam in cell $i$ and $p_e(\bm{h}_i)$ is the routing probability. This loss aligns expert assignment with the dominant propagation direction.

\emph{Missing-beam penalty} ($\mathcal{L}_{\mathrm{reg}}$) addresses beams absent from raw MRs due to reporting filters or beam failures. Because these entries lack ground-truth RSRP, we impose a hinge penalty to prevent implausibly strong predictions:
\begin{equation}
\mathcal{L}_{\mathrm{reg}}\;=\!\!\!\sum_{m\in\mathcal{M}}\!\!\!
\max\!\bigl(0,\;\hat{y}_m - y_{\min}\bigr),
\end{equation}
where $\mathcal{M}$ is the set of originally missing beams and $y_{\min}$ is the minimum reported RSRP among the cell's visible beams.

\paragraph{Multi-granularity masking strategy.}
At inference time, \charT{} must handle both beam-level queries (reconstructing individual missing beams within an observed cell) and cell-level queries (predicting coverage for an entirely unobserved cell). To ensure that the model encounters both granularities during training---a necessary condition for learning to predict at both levels\cite{zhou2026incomplete}---we adopt a multi-granularity masking strategy: with probability $p_{\mathrm{intra}}=0.4$, a sample receives \emph{beam-level} masking where a random subset of beams within each cell is hidden; with probability $1-p_{\mathrm{intra}}$, the sample receives \emph{cell-level} masking where all beams of randomly chosen cells are hidden (\figref{fig:method}\textbf{d}, left). Mask ratios are sampled uniformly from 20\%--80\%, exposing the model to the full range of sparsity conditions encountered during deployment.

\paragraph{Channel-model-constrained augmentation.}
Effective augmentation for the teacher's view $\mathcal{S}'$ requires identifying and perturbing physically meaningful parameters---transmission power, beam codebook, antenna orientation---while preserving the structural invariants of radio propagation. The channel model in \eqref{eq:lscm} provides the key decomposition: it factors each observation into a \emph{configuration-dependent} component (the beamforming pattern matrix $\bm{A}$ and the transmission power) and an \emph{environment-dependent} component (the channel APS $\bm{x}$), with their linear mapping $\bm{y} = \bm{A}\bm{x}$ serving as the propagation invariant that any valid augmentation must respect. Na\"ive perturbations that ignore this structure, such as additive noise or random beam dropout, can produce observation--configuration pairs that violate the invariant and corrupt the alignment target.

In the \emph{power domain}, a multiplicative factor sampled uniformly from $[a,\,b]$ and additive Gaussian noise are applied jointly to RSRP values and the corresponding transmission power in $\bm{c}_{i,j}$, perturbing the power-related parameter while preserving the linear RSRP--power relationship dictated by \eqref{eq:lscm}. In the \emph{beam domain}, the channel APS $\hat{\bm{x}}$---the propagation invariant---is first extracted from observed beams via sparse recovery on \eqref{eq:lscm}, isolating the environment-dependent structure from the incomplete observations. Given this estimated APS, augmented views are generated by varying the configuration-dependent component: the beam codebook or antenna orientation is perturbed to produce a modified pattern matrix $\tilde{\bm{A}}$, and a new beam vector $\tilde{\bm{y}} = \tilde{\bm{A}}\hat{\bm{x}}$ is synthesized through the forward model (\figref{fig:method}\textbf{d}, left). This procedure simultaneously densifies missing entries and exposes the model to configurations not present in the original data, exposing the model to configuration-dependent RSRP variations under the channel-model assumption that the estimated propagation component remains fixed. Cells with severely incomplete observations are excluded to reduce instability in the APS estimation used to construct the teacher’s augmented view. By constraining all perturbations to respect the propagation invariant in \eqref{eq:lscm}, the augmented views expose the model to diverse power and configuration conditions while ensuring that the learned representations remain physically grounded.

%% ============================================================
\subsection*{Applications for network intelligence tasks}

The MBM formulation in \eqref{eq:mbm} supports two forms of transfer. Tasks with beam-level RSRP outputs are formulated directly as context-conditioned queries to the pretrained predictor, whereas tasks requiring other outputs reuse the frozen representation through lightweight task-specific heads.

\paragraph{Zero-shot inference.}
For radio map construction, beams with missing RSRP constitute the target set $\mathcal{S}_t$, while the remaining observations form the context set $\mathcal{S}_c$. For new-site deployment prediction, all beams of a planned cell are treated as target queries and observations from surrounding cells provide the network context. For parameter tuning prediction, the target cell is queried using the proposed antenna configuration, while observations from the remaining network form $\mathcal{S}_c$. In all three tasks, the pretrained predictor directly estimates the target beam-level RSRP without gradient updates or target-city calibration.

\paragraph{Few-shot adaptation.}
For applications requiring outputs beyond beam-level RSRP, the pretrained backbone $\mathcal{E}_{\theta}$ remains frozen and only lightweight task-specific heads are adapted using a small labelled subset from the target city. A source-city-trained head provides the initialization, enabling efficient cross-city adaptation without updating the shared representation. UE localization and SINR estimation use MLP regression heads optimized with mean Euclidean distance and the $L_1$ loss, respectively. Beam prediction and propagation scenario classification use MLP classifiers with softmax outputs and are optimized using cross-entropy.

Beyond these prediction tasks, \charT{} supports beamspace clustering for organizing large-scale MR data. An MLP embedding head maps the pooled cell-level representations into a sample-level beamspace, where deep embedded clustering jointly refines the embeddings and learnable grid prototypes\cite{xie2016unsupervised}. Prototype-based assignment discretizes the continuous radio-data space into reusable radio grids for downstream network modelling and analysis\cite{wang2026learning}.

\section*{Data availability}
A subset of the data used in this study will be made publicly
available at \url{https://github.com/rorschaches/RadioWorld}.

\section*{Code availability}
Once the paper is accepted, the code will be open-sourced via GitHub at \url{https://github.com/xyqin-cuhk/ChaRT}.

\section*{Funding}
The work was supported in part by the National Key Research and
Development Program of China under Grant 2024YFA1014201 awarded to Tsung-Hui Chang.

\section*{Ethics declaration}
As described in the Methods, all user equipment
identifiers were removed. A location-tagged subset was obtained
through a crowdsourcing programme in which participating users
consented to the collection of GPS coordinates alongside signal
measurements. Anonymized coordinates were used for localization
and radio-map visualization.

\section*{Author contributions}
W.P., T.-H.C. and Z.-Q.L. conceived and planned the study.
X.Q. and T.-H.C. developed the methodology, and X.Q. developed
the model. X.Q. cleaned the data, with additional
data curation by L.S. and Q.Y. X.Q. and H.D. conducted the
experiments, and H.D. prepared the figures. B.P., Y.X., L.S.,
Q.Y. and Z.-Q.L. analysed the results. X.Q. wrote the initial
manuscript. W.P. and T.-H.C. supervised the research.
All authors reviewed and edited the manuscript and approved
the submitted version.

\section*{Competing interests}
The authors declare no competing interests.

% =====================================================================
%  Bibliography (external file: refs.bib)
% =====================================================================
\bibliographystyle{unsrtnat}
\bibliography{refs}

@techreport{GSMA2026MobileEconomy,
  author      = {{GSMA}},
  title       = {The Mobile Economy 2026},
  institution = {GSMA},
  year        = {2026},
  url         = {https://www.gsma.com/solutions-and-impact/connectivity-for-good/mobile-economy/wp-content/uploads/2026/02/The-Mobile-Economy-2026.pdf}
}

@article{Li2022RealWorld,
  author  = {Li, Yang and Zhang, Shutao and Ren, Xiaohui and Zhu, Jianhang and Huang, Jiajie and He, Pengcheng and Shen, Kaiming and Yao, Zhiqiang and Gong, Jie and Chang, Tsung-Hui and Shi, Qingjiang and Luo, Zhi-Quan},
  title   = {Real-World Wireless Network Modeling and Optimization: From Model/Data-Driven Perspective},
  journal = {Chinese Journal of Electronics},
  volume  = {31},
  pages   = {991--1012},
  year    = {2022},
  doi     = {10.1049/cje.2022.00.191}
}

@article{Khan2022,
  author  = {Khan, Latif U. and Saad, Walid and Niyato, Dusit and Han, Zhu and Hong, Choong Seon},
  title   = {Digital-Twin-Enabled {6G}: Vision, Architectural Trends, and Future Directions},
  journal = {IEEE Communications Magazine},
  year    = {2022},
  volume  = {60},
  number  = {1},
  pages   = {74--80},
  doi     = {10.1109/MCOM.001.21143}
}

@techreport{ITUT2022,
  author      = {{ITU-T}},
  title       = {Digital Twin Network -- Requirements and Architecture},
  institution = {International Telecommunication Union},
  year        = {2022},
  number      = {Recommendation Y.3090},
  url         = {https://www.itu.int/rec/T-REC-Y.3090-202202-I}
}

@article{Mihai2022DTSurvey,
  author  = {Mihai, Stefan and Yaqoob, Mahnoor and Hung, Dang V. and Davis, William and Towakel, Panos and Raza, Mohsen and Karamanoglu, Mehmet and Barn, Balbir and Shetve, Dhaval and Prasad, R. Venkatesha and Venkataraman, Hrishikesh and Trestian, Ramona and Nguyen, Huan X.},
  title   = {Digital Twins: A Survey on Enabling Technologies, Challenges, Trends and Future Prospects},
  journal = {IEEE Communications Surveys \& Tutorials},
  year    = {2022},
  volume  = {24},
  number  = {4},
  pages   = {2255--2291},
  doi     = {10.1109/COMST.2022.3208773}
}

@article{HuiDTN2023,
  author  = {Hui, Linbo and Wang, Mowei and Zhang, Liang and Lu, Lu and Cui, Yong},
  title   = {Digital Twin for Networking: A Data-Driven Performance Modeling Perspective},
  journal = {IEEE Network},
  year    = {2023},
  volume  = {37},
  number  = {3},
  pages   = {200--208}
}

@techreport{3GPP_38.331,
  title     = {{NR; Radio Resource Control (RRC); Protocol specification}},
  author    = {{3GPP}},
  institution = {{3rd Generation Partnership Project (3GPP)}},
  year      = {2025},
  type      = {Technical Specification (TS)},
  number    = {38.331},
  note      = {Version 16.20.0},
   url      = {https://portal.3gpp.org/desktopmodules/Specifications/SpecificationDetails.aspx?specificationId=3197}
}

@techreport{3GPP_38.901,
	author = {{3GPP}},
	institution = {3rd Generation Partnership Project (3GPP)},
	number = {38.901},
	note = {Version 16.1.0},
	title = {Study on channel model for frequencies from 0.5 to 100 {GHz}},
	type = {Technical Report (TR)},
	year = {2020},
	month = {Jan.},
        url = {https://portal.3gpp.org/desktopmodules/Specifications/SpecificationDetails.aspx?specificationId=3173}
    }

@inproceedings{Brown2020,
  author    = {Brown, Tom B. and Mann, Benjamin and Ryder, Nick and Subbiah, Melanie and Kaplan, Jared and
               Dhariwal, Prafulla and Neelakantan, Arvind and Shyam, Pranav and Sastry, Girish and Askell, Amanda and
               Agarwal, Sandhini and Herbert-Voss, Ariel and Krueger, Gretchen and Henighan, Tom and Child, Rewon and
               Ramesh, Aditya and Ziegler, Daniel M. and Wu, Jeffrey and Winter, Clemens and Hesse, Christopher and
               Chen, Mark and Sigler, Eric and Litwin, Mateusz and Gray, Scott and Chess, Benjamin and Clark, Jack and
               Berner, Christopher and McCandlish, Sam and Radford, Alec and Sutskever, Ilya and Amodei, Dario},
  title     = {Language Models are Few-Shot Learners},
  booktitle = {Advances in Neural Information Processing Systems (NeurIPS)},
  volume    = {33},
  pages     = {1877--1901},
  year      = {2020}
}

@article{Touvron2023,
  author  = {Touvron, Hugo and Lavril, Thibaut and Izacard, Gautier and Martinet, Xavier and Lachaux, Marie-Anne and
             Lacroix, Timoth{\'e}e and Rozi{\`e}re, Baptiste and Goyal, Naman and Hambro, Eric and Azhar, Faisal and
             Rodriguez, Aurelien and Joulin, Armand and Grave, Edouard and Lample, Guillaume},
  title   = {{LLaMA}: Open and Efficient Foundation Language Models},
  journal = {arXiv preprint arXiv:2302.13971},
  year    = {2023},
  url     = {https://arxiv.org/abs/2302.13971}
}

@inproceedings{Radford2021,
  author    = {Radford, Alec and Kim, Jong Wook and Hallacy, Chris and Ramesh, Aditya and Goh, Gabriel and
               Agarwal, Sandhini and Sastry, Girish and Askell, Amanda and Mishkin, Pamela and Clark, Jack and
               Krueger, Gretchen and Sutskever, Ilya},
  title     = {Learning Transferable Visual Models from Natural Language Supervision},
  booktitle = {Proceedings of the 38th International Conference on Machine Learning (ICML)},
  year      = {2021},
  pages     = {8748--8763}
}

@InProceedings{Kirillov2023,
    author    = {Kirillov, Alexander and Mintun, Eric and Ravi, Nikhila and Mao, Hanzi and Rolland, Chloe and Gustafson, Laura and Xiao, Tete and Whitehead, Spencer and Berg, Alexander C. and Lo, Wan-Yen and Dollar, Piotr and Girshick, Ross},
    title     = {Segment Anything},
    booktitle = {Proceedings of the IEEE/CVF International Conference on Computer Vision (ICCV)},
    month     = {October},
    year      = {2023},
    pages     = {4015--4026}
}

@InProceedings{Caron2021,
    author    = {Caron, Mathilde and Touvron, Hugo and Misra, Ishan and J\'egou, Herv\'e and Mairal, Julien and Bojanowski, Piotr and Joulin, Armand},
    title     = {Emerging Properties in Self-Supervised Vision Transformers},
    booktitle = {Proceedings of the IEEE/CVF International Conference on Computer Vision (ICCV)},
    month     = {October},
    year      = {2021},
    pages     = {9650--9660}
}

@article{SkySensePP2025,
  author  = {Wu, Kang and Zhang, Yingying and Ru, Lixiang and Dang, Bo and
             Lao, Jiangwei and Yu, Lei and Luo, Junwei and Zhu, Zifan and
             Sun, Yue and Zhang, Jiahao and Zhu, Qi and Wang, Jian and
             Yang, Ming and Chen, Jingdong and Zhang, Yongjun and Li, Yansheng},
  title   = {A semantic-enhanced multi-modal remote sensing foundation model
             for {Earth} observation},
  journal = {Nature Machine Intelligence},
  year    = {2025},
  volume  = {7},
  pages   = {1235--1249},
  doi     = {10.1038/s42256-025-01078-8}
}

@article{CSFM2026,
  author  = {Gu, Xiao and Tang, Wei and Han, Jinpei and Sangha, Veer and
             Liu, Fenglin and Gowda, Shreyank N. and Ribeiro, Antonio H. and
             Schwab, Patrick and Branson, Kim and Clifton, Lei and
             Ribeiro, Antonio Luiz P. and Liu, Zhangdaihong and
             Clifton, David A.},
  title   = {Cardiac health assessment across scenarios and devices using
             a multimodal foundation model pretrained on data from
             1.7 million individuals},
  journal = {Nature Machine Intelligence},
  year    = {2026},
  volume  = {8},
  pages   = {220--233},
  doi     = {10.1038/s42256-026-01180-5}
}

@article{LucaOne2025,
  author  = {He, Yong and Fang, Pan and Shan, Yongtao and Pan, Yuanfei and Wei, Yanhong and Chen, Yichang and
             Chen, Yihao and Liu, Yi and Zeng, Zhenyu and Zhou, Zhan and Zhang, Feng and Esmaeili, Adibvafa and
             Bao, Yongchao and Zhang, Le and Tao, Aoyun and Sun, Hao and Yang, Tao and Sun, Liyang and
             Wang, Sheng and Zhao, Tongqing and Cui, Yuhao and Han, Lin},
  title   = {Generalized biological foundation model with unified nucleic acid and protein language},
  journal = {Nature Machine Intelligence},
  year    = {2025},
  doi     = {10.1038/s42256-025-01044-4}
}

@article{wifo2,
    author = {Liu, Boxun and Liu, Xuanyu and Gao, Shijian and Cai, Xuesong and Cheng, Xiang and Yang, Liuqing},
    title = {WiFo-2: a generalist foundation model unifies heterogeneous wireless system design},
    journal = {National Science Review},
    volume = {13},
    number = {18},
    pages = {nwag500},
    year = {2026},
    month = {09},
    issn = {2095-5138},
    doi = {10.1093/nsr/nwag500},
    url = {https://doi.org/10.1093/nsr/nwag500},
    eprint = {https://academic.oup.com/nsr/article-pdf/13/18/nwag500/70672408/nwag500.pdf},
}

@article{SpectrumFM2025,
  author  = {Liu, Fuhui and Zhang, Hangyu and Wang, Wei and Zhang, Wei and Wang, Cheng-Xiang and Niyato, Dusit and
             Cui, Shuguang},
  title   = {{SpectrumFM}: A Foundation Model for Intelligent Spectrum Management},
  journal = {arXiv preprint arXiv:2505.06256},
  year    = {2025},
  url     = {https://arxiv.org/abs/2505.06256}
}

@article{ICWLM2025,
  author  = {Wen, Yuxuan and Cui, Yi and Zhang, Yifei and Yu, Wei and Han, Zhu and Zhao, Tony Q. S.},
  title   = {{ICWLM}: A Multi-Task Wireless Large Model via In-Context Learning},
  journal = {arXiv preprint arXiv:2507.18167},
  year    = {2025},
  url     = {https://arxiv.org/abs/2507.18167}
}

@article{LLM4WM2025,
  author  = {Liu, Xinyu and Gao, Xiaoyan and Liu, Boxun and Cheng, Xinyu and Yang, Lei},
  title   = {{LLM4WM}: Adapting {LLM} for Wireless Multi-Tasking},
  journal = {arXiv preprint arXiv:2501.12983},
  year    = {2025},
  url     = {https://arxiv.org/abs/2501.12983}
}

@article{AlikhateebLWM2024,
  author  = {Alikhateeb, Ahmed and Charan, Gokul and Alkhateeb, Tareq},
  title   = {Large Wireless Model ({LWM}): A Foundation Model for Wireless Channels},
  journal = {arXiv preprint arXiv:2411.08872},
  year    = {2024},
  url     = {https://arxiv.org/abs/2411.08872}
}

@article{CatakBERT4MIMO2025,
  author  = {Catak, Ferhat Ozgur and Kuzlu, Murat and Cali, Ulrich},
  title   = {{BERT4MIMO}: A Foundation Model Using {BERT} Architecture for Massive {MIMO} Channel State Information Prediction},
  journal = {arXiv preprint arXiv:2501.01802},
  year    = {2025},
  url     = {https://arxiv.org/abs/2501.01802}
}

@article{QuaDRiGa2014,
  author  = {Jaeckel, Stephan and Raschkowski, Leszek and B{\"o}rner, Karl and Thiele, Lars},
  title   = {{QuaDRiGa}: A 3-D Multi-Cell Channel Model With Time Evolution for Enabling Virtual Field Trials},
  journal = {IEEE Transactions on Antennas and Propagation},
  year    = {2014},
  volume  = {62},
  number  = {6},
  pages   = {3242--3256},
  doi     = {10.1109/TAP.2014.2310222}
}

@article{SalihuLocalization2024,
  author  = {Salihu, Abubakar and Rupp, Markus and Schwarz, Stefan},
  title   = {Self-Supervised and Invariant Representations for Wireless Localization},
  journal = {IEEE Transactions on Wireless Communications},
  year    = {2024},
  volume  = {23},
  number  = {8},
  pages   = {8281--8296},
  doi     = {10.1109/TWC.2023.3348203}
}

@article{hu2024localization,
  author  = {Hu, Zhinan and Chen, Xin and Zhou, Zhenyu and Mumtaz, Shahid},
  title   = {Localization with cellular signal {RSRP} fingerprint of multiband and multicell},
  journal = {IEEE Journal on Selected Areas in Communications},
  year    = {2024},
  volume  = {42},
  number  = {9},
  pages   = {2380--2394},
  doi     = {10.1109/JSAC.2024.3414000}
}

@article{Qin2026MRLSCM,
  author  = {Qin, Xinyu and Yan, Qi and Zhang, Shutao and Peng, Bingsheng and
             Xue, Ye and Chang, Tsung-Hui},
  title   = {A Measurement Report Data-Driven Framework for Localized
             Statistical Channel Modeling},
  journal = {IEEE Transactions on Mobile Computing},
  year    = {2026},
  pages   = {1--15},
  doi     = {10.1109/TMC.2026.3667749}
}

@article{AlrabeiahBeam2020,
  author  = {Alrabeiah, Mohammed and Alkhateeb, Ahmed},
  title   = {Deep Learning for mmWave Beam and Blockage Prediction Using Sub-6 GHz Channels},
  journal = {IEEE Transactions on Communications},
  year    = {2020},
  volume  = {68},
  number  = {9},
  pages   = {5504--5518},
  doi     = {10.1109/TCOMM.2020.3006325}
}

@inproceedings{Schmitz2012SphericalHarmonics,
  author    = {Schmitz, Arne and Karolski, Thomas and Kobbelt, Leif},
  title     = {Using Spherical Harmonics for Modeling Antenna Patterns},
  booktitle = {2012 IEEE Radio and Wireless Symposium},
  pages     = {155--158},
  year      = {2012},
  publisher = {IEEE},
  doi       = {10.1109/RWS.2012.6175298},
  url       = {https://doi.org/10.1109/RWS.2012.6175298}
}

@misc{Sionna2025,
      title={Sionna {RT}: Technical Report},
      author={Ait Aoudia, Faycal and Hoydis, Jakob and Nimier-David, Merlin and Nicolet, Baptiste and Cammerer, Sebastian and Keller, Alexander},
      year={2025},
      eprint={2504.21719},
      archivePrefix={arXiv},
      primaryClass={cs.IT},
      url={https://arxiv.org/abs/2504.21719},
}

@misc{Remcom2023,
  author       = {{Remcom}},
  title        = {{Wireless InSite}: Site-Specific Radio Propagation Modelling Software},
  year         = {2023},
  howpublished = {\url{https://www.remcom.com/wireless-insite-em-propagation-software}},
  note         = {Accessed: 2026-05-02}
}

@inproceedings{xgboost,
  author    = {Chen, Tianqi and Guestrin, Carlos},
  title     = {{XGBoost}: A Scalable Tree Boosting System},
  booktitle = {Proceedings of the 22nd ACM SIGKDD International Conference
               on Knowledge Discovery and Data Mining (KDD)},
  year      = {2016},
  pages     = {785--794},
  doi       = {10.1145/2939672.2939785}
}

@article{tabpfn_v2,
  author  = {Hollmann, Noah and M{\"u}ller, Samuel and Purucker, Lennart and
             Krishnakumar, Arjun and K{\"o}rfer, Max and Hoo, Shi Bin and
             Schirrmeister, Robin Tibor and Hutter, Frank},
  title   = {Accurate predictions on small data with a tabular foundation model},
  journal = {Nature},
  year    = {2025},
  volume  = {637},
  pages   = {319--326},
  doi     = {10.1038/s41586-024-08328-6}
}

@article{lscm,
  author  = {Zhang, Shutao and Ning, Xinzhi and Zheng, Xi and Shi, Qingjiang
             and Chang, Tsung-Hui and Luo, Zhi-Quan},
  title   = {A Physics-Based and Data-Driven Approach for Localized
             Statistical Channel Modeling},
  journal = {IEEE Transactions on Wireless Communications},
  year    = {2024},
  volume  = {23},
  number  = {6},
  pages   = {5409--5424},
  doi     = {10.1109/TWC.2023.3326209}
}

@inproceedings{zhou2026incomplete,
    title={Incomplete Data, Complete Dynamics: A Diffusion Approach},
    author={Zihan Zhou and Chenguang Wang and Hongyi Ye and Yongtao Guan and Tianshu Yu},
    booktitle={The Fourteenth International Conference on Learning Representations},
    year={2026},
    url={https://openreview.net/forum?id=NYvvkBlSX2}
}

@InProceedings{He2022MAE,
    author    = {He, Kaiming and Chen, Xinlei and Xie, Saining and Li, Yanghao and Doll\'ar, Piotr and Girshick, Ross},
    title     = {Masked Autoencoders Are Scalable Vision Learners},
    booktitle = {Proceedings of the IEEE/CVF Conference on Computer Vision and Pattern Recognition (CVPR)},
    month     = {June},
    year      = {2022},
    pages     = {16000-16009}
}

@InProceedings{Lee2019SetTransformer,
  title = 	 {Set Transformer: A Framework for Attention-based Permutation-Invariant Neural Networks},
  author =       {Lee, Juho and Lee, Yoonho and Kim, Jungtaek and Kosiorek, Adam and Choi, Seungjin and Teh, Yee Whye},
  booktitle = 	 {Proceedings of the 36th International Conference on Machine Learning},
  pages = 	 {3744--3753},
  year = 	 {2019},
  editor = 	 {Chaudhuri, Kamalika and Salakhutdinov, Ruslan},
  volume = 	 {97},
  series = 	 {Proceedings of Machine Learning Research},
  month = 	 {09--15 Jun},
  publisher =    {PMLR},
}

@article{wang2026learning,
  title={Learning to Gridize: Segment Physical World by Wireless Communication Channel},
  author={Wang, Juntao and Yin, Feng and Ding, Tian and Chang, Tsung-Hui and Luo, Zhi-Quan and Yan, Qi},
  journal={IEEE Transactions on Mobile Computing},
  year={2026},
  publisher={IEEE}
}

@inproceedings{xie2016unsupervised,
  title={Unsupervised deep embedding for clustering analysis},
  author={Xie, Junyuan and Girshick, Ross and Farhadi, Ali},
  booktitle={International conference on machine learning},
  pages={478--487},
  year={2016},
  organization={PMLR}
}

@inproceedings{Dai2024DeepSeekMoE,
  title     = {{D}eep{S}eek{M}o{E}: Towards Ultimate Expert Specialization in Mixture-of-Experts Language Models},
  author    = {Dai, Damai and Deng, Chengqi and Zhao, Chenggang and Xu, R. X. and Gao, Huazuo and Chen, Deli and Li, Jiashi and Zeng, Wangding and Yu, Xingkai and Wu, Y. and Xie, Zhenda and Li, Y. K. and Huang, Panpan and Luo, Fuli and Ruan, Chong and Sui, Zhifang and Liang, Wenfeng},
  booktitle = {Proceedings of the 62nd Annual Meeting of the Association for Computational Linguistics (Volume 1: Long Papers)},
  pages     = {1280--1297},
  year      = {2024},
  month     = aug,
  address   = {Bangkok, Thailand},
  publisher = {Association for Computational Linguistics},
  doi       = {10.18653/v1/2024.acl-long.70},
  url       = {https://aclanthology.org/2024.acl-long.70/}
}

% =====================================================================
%  Extended Data
% =====================================================================
\clearpage
\section*{Extended Data}

\setcounter{figure}{0}
\renewcommand{\thefigure}{\arabic{figure}}
\renewcommand{\theHfigure}{ED.\arabic{figure}}
\setcounter{table}{0}
\renewcommand{\thetable}{\arabic{table}}
\renewcommand{\theHtable}{ED.\arabic{table}}
\captionsetup[figure]{name={Extended Data Fig.},labelfont=bf,labelsep=naturebar,justification=raggedright,singlelinecheck=false}

% % =====================================================================
% % Extended Data Table 1
% % =====================================================================

% \begin{table}[!htbp]
% \refstepcounter{table}
% \label{tab:ed1}
% \noindent\textbf{Extended Data Table~\thetable\ \textbar{} Per-city dataset statistics.}\par
% \vspace{6pt}
% \centering
% \renewcommand{\arraystretch}{1.15}
% \begin{tabular}{@{}lrrrcc@{}}
% \toprule
% City & MR samples & Beam observations & Cells & Area (km$^2$) & Carrier (GHz) \\
% \midrule
% City A & 1{,}003{,}362{,}798 & 18{,}207{,}030{,}406 & 3{,}503 & $16\!\times\!18$ & 2.1\,/\,3.5 \\
% City B & 598{,}023 & 8{,}795{,}040 & 280 & $8\!\times\!7$ & 2.6 \\
% City C & 702{,}732 & 6{,}382{,}334 & 16 & $2.5\!\times\!2.5$ & 3.5 \\
% \midrule
% \textbf{Total} & \textbf{1{,}004{,}663{,}553} & \textbf{18{,}222{,}207{,}780} & \textbf{3{,}799} & -- & -- \\
% \bottomrule
% \end{tabular}
% \par\vspace{3pt}
% \begin{minipage}{0.95\linewidth}
% \footnotesize\raggedright
% \textit{Note:} MR samples denote individual measurement reports, and beam observations denote beam-level RSRP records. City~A is a dense multi-band urban deployment, City~B is a moderately dense single-band urban deployment and City~C is a sparse single-band suburban deployment. Bold values denote dataset totals.
% \end{minipage}
% \end{table}

% =====================================================================
% Extended Data Table 2
% =====================================================================

\begin{table}[!htbp]
\refstepcounter{table}
\label{tab:ed2}
\noindent\textbf{Extended Data Table~\thetable\ \textbar{} Parameter changes across collection rounds.}\par
\vspace{6pt}
\centering
\renewcommand{\arraystretch}{1.2}
\begin{tabular}{@{}lcc@{}}
\toprule
\textbf{Parameter type} & \textbf{Round~1 $\rightarrow$ Round~2} & \textbf{Round~1 $\rightarrow$ Round~3} \\
\midrule
Mechanical azimuth & 3 & 12 \\
Mechanical tilt & 8 & 15 \\
Transmit power & 2 & 4 \\
Electrical tilt & 3 & 19 \\
Beam codebook & 2 & 11 \\
\bottomrule
\end{tabular}
\par\vspace{3pt}
\begin{minipage}{0.90\linewidth}
\footnotesize\raggedright
\textit{Note:} Round~1 defines the reference antenna configuration. Entries report the number of cells for which each parameter changed in Round~2 or Round~3 relative to Round~1. For the parameter-tuning evaluation, the affected cells are masked in their entirety, and the frozen \charT{} model predicts their post-tuning multi-beam coverage from the surrounding network context and updated configuration vectors.
\end{minipage}
\end{table}

% =====================================================================
% Extended Data Table 3
% =====================================================================

\begin{table}[!htbp]
\refstepcounter{table}
\label{tab:ed3_ablation}
\noindent\textbf{Extended Data Table~\thetable\ \textbar{} Ablation of \charT{} components.}\par
\vspace{6pt}
\centering
\renewcommand{\arraystretch}{1.2}
\begin{tabular}{@{}lc@{}}
\toprule
\textbf{Model configuration} & \textbf{MAE (dB)} \\
\midrule
Contextual self-distillation & 6.14 \\
$+$ Channel-model-constrained augmentation & 5.71 \\
$+$ Hierarchical masked attention & 4.98 \\
$+$ CSG-MoE (\charT{}) & \textbf{4.72} \\
\bottomrule
\end{tabular}
\par\vspace{3pt}
\begin{minipage}{0.90\linewidth}
\footnotesize\raggedright
\textit{Note:} Components are added cumulatively to the preceding configuration. All variants are evaluated using cell-level RSRP reconstruction at a 40\% masking ratio under spatial extrapolation in City~A. MAE denotes mean absolute error. Lower values indicate better performance, and bold denotes the best result.
\end{minipage}
\end{table}

% =====================================================================
% Extended Data Figures
% =====================================================================

\clearpage
\begin{figure*}[p]
\centering
\includegraphics[width=\textwidth]{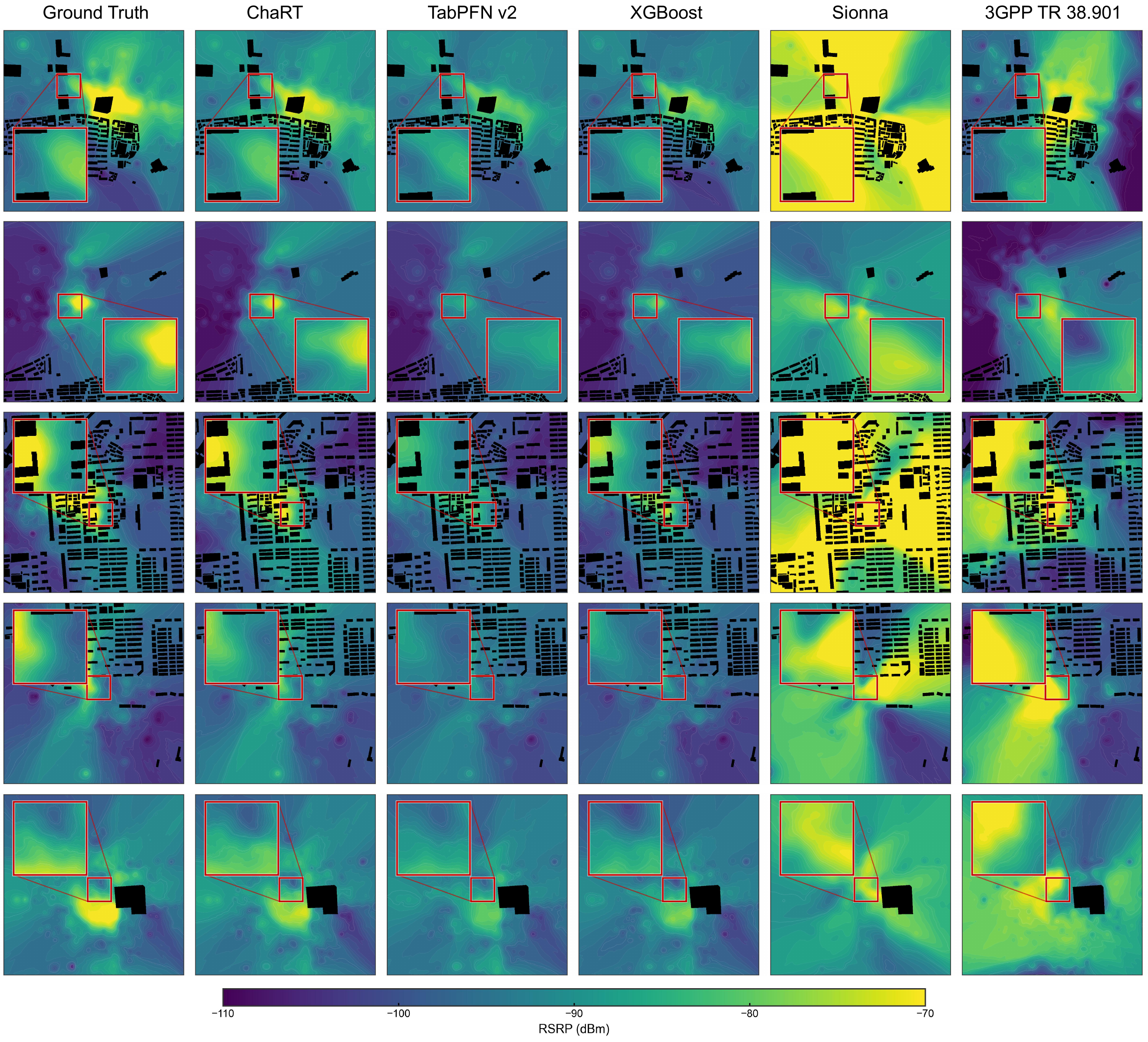}
\caption{\textbf{Radio map comparison for cross-city zero-shot new-site deployment in City~B.} Each row corresponds to one target cell located at the centre of the map. Building footprints are overlaid to indicate the surrounding urban environment. Columns show the measured RSRP and predictions from \charT{}, TabPFN~v2, XGBoost, Sionna and the 3GPP TR~38.901 model. RSRP is shown in dBm, and the insets magnify regions exhibiting pronounced spatial variation. The anonymized UE coordinates used to render the maps are obtained from a sparsely location-tagged subset of MR samples collected through a consenting-user crowdsourcing programme\cite{Qin2026MRLSCM}. Measurements and predictions are spatially interpolated only for visualization.}
\label{fig:radiomap_new_bs}
\end{figure*}

\clearpage
\begin{figure*}[p]
\centering
\includegraphics[width=\textwidth]{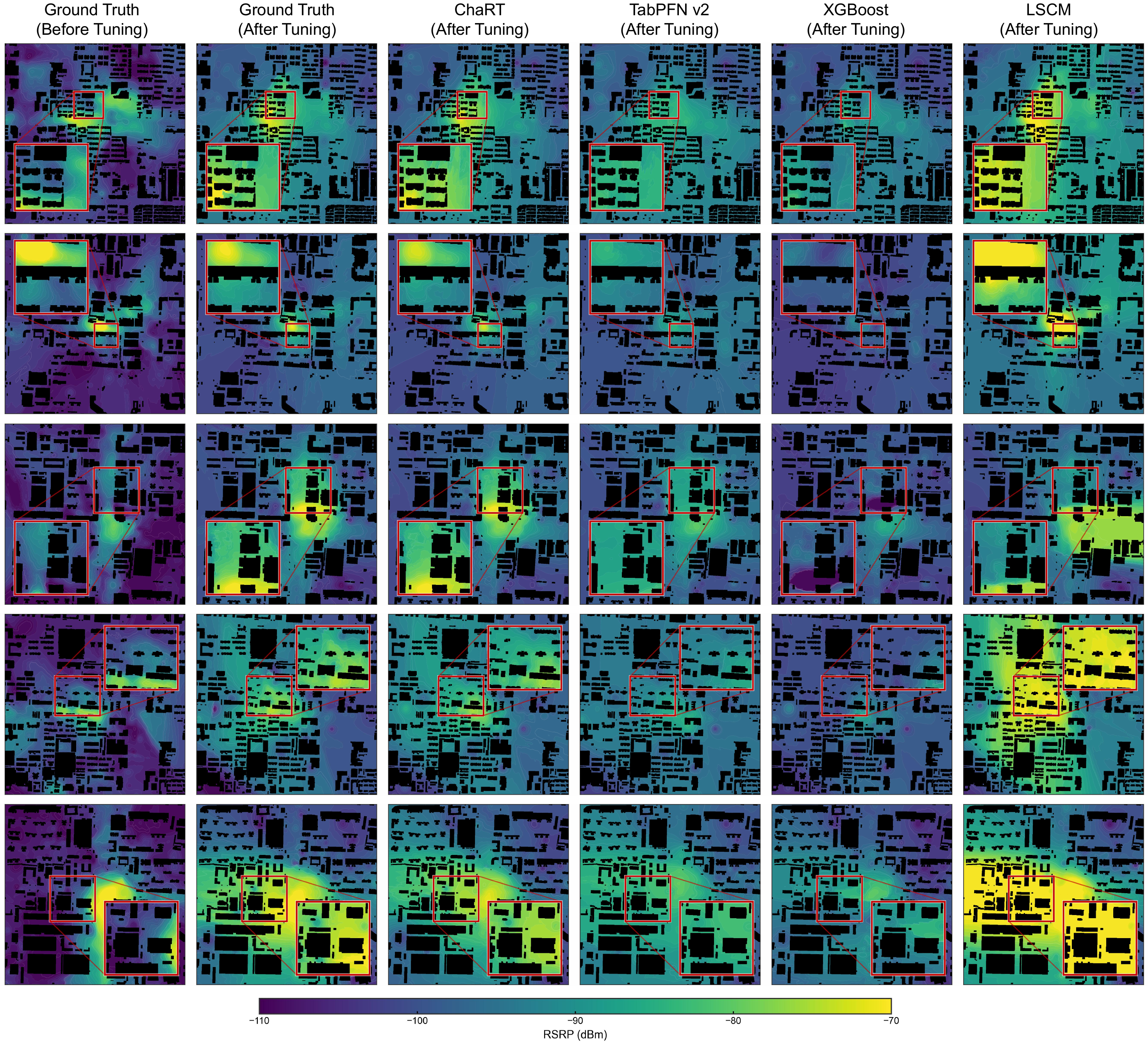}
\caption{\textbf{Radio map comparison under antenna parameter tuning.} Each row presents a representative cell for which one antenna parameter was adjusted. From top to bottom, the changes concern mechanical azimuth, mechanical tilt, transmit power, electrical tilt and beam codebook. The corresponding settings change from $0^\circ$ to $350^\circ$, $4^\circ$ to $7^\circ$, $-0.2534$ to $1.747$\,dBm per resource element, $3^\circ$ to $6^\circ$, and beam-codebook index 0 to 7, respectively. The target cell is located at the centre of each map, and building footprints indicate the surrounding urban environment. Columns show the measured RSRP before and after tuning and the post-tuning predictions from \charT{}, TabPFN~v2, XGBoost and LSCM. The anonymized UE coordinates used to render the maps are obtained from a sparsely location-tagged subset of MR samples collected through a consenting-user crowdsourcing programme\cite{Qin2026MRLSCM}. Measurements and predictions are spatially interpolated only for visualization.}
\label{fig:radiomap_tuning}
\end{figure*}

\clearpage
\begin{figure*}[p]
\centering
\includegraphics[width=\textwidth]{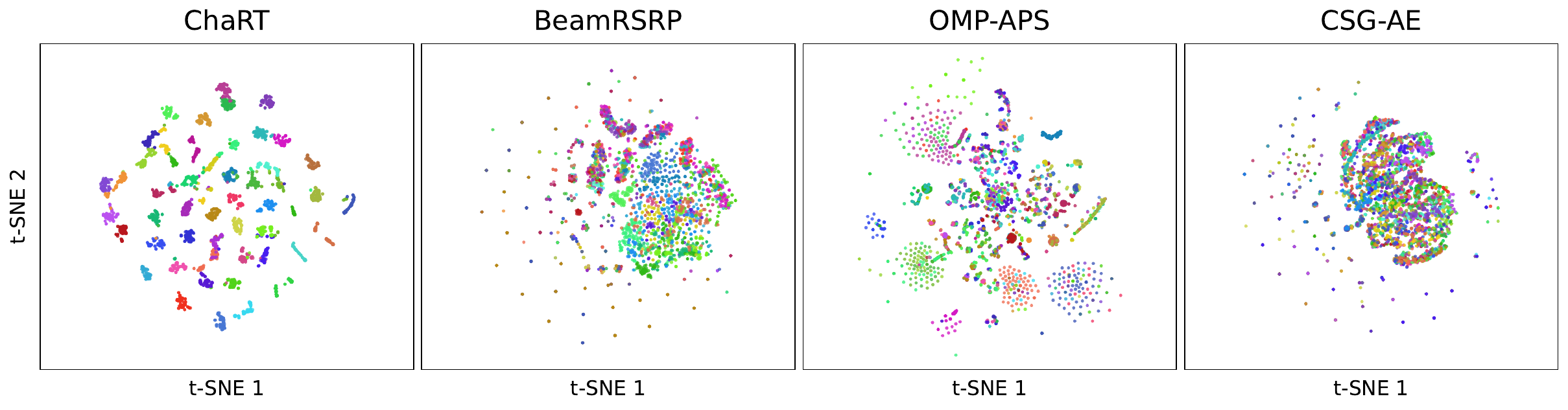}
\caption{\textbf{Latent organization of beamspace-clustering representations on the held-out Round~3 data.} t-SNE projections are shown for 50 representative radio grids from each method, with colours denoting grid assignments. The projections provide a qualitative comparison of inter-grid mixing and the organization of the learned representation spaces; they are not used to quantify cluster separation or downstream performance.}
\label{fig:cluster_tsne}
\end{figure*}

\clearpage
\begin{figure*}[p]
\centering
\begin{subfigure}[t]{0.485\linewidth}
\phantomcaption
\label{fig:ed_csg_moe_rsrp}
\raggedright\textbf{a}\par\vspace{-1pt}
\includegraphics[width=\linewidth]{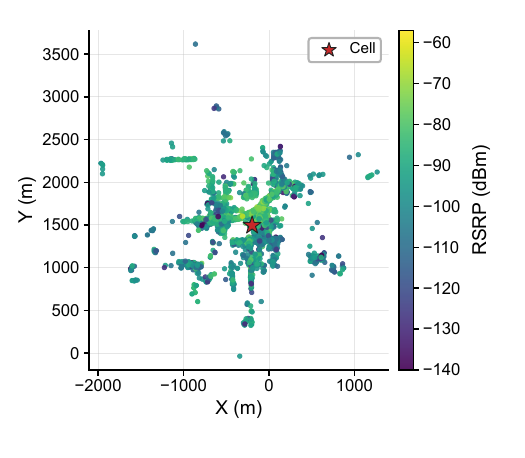}
\end{subfigure}
\hfill
\begin{subfigure}[t]{0.485\linewidth}
\phantomcaption
\label{fig:ed_csg_moe_beam}
\raggedright\textbf{b}\par\vspace{-1pt}
\includegraphics[width=\linewidth]{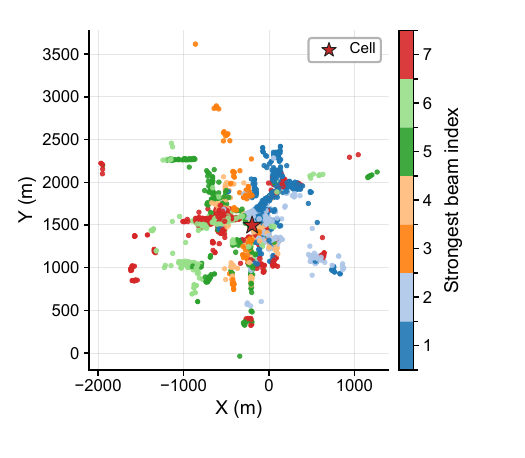}
\end{subfigure}

\par\vspace{-0.3em}

\begin{subfigure}[t]{0.485\linewidth}
\phantomcaption
\label{fig:ed_vanilla_moe}
\raggedright\textbf{c}\par\vspace{-1pt}
\includegraphics[width=\linewidth]{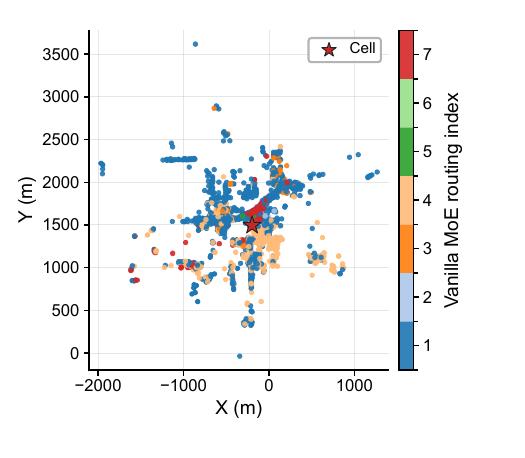}
\end{subfigure}
\hfill
\begin{subfigure}[t]{0.485\linewidth}
\phantomcaption
\label{fig:ed_csg_moe_routing}
\raggedright\textbf{d}\par\vspace{-1pt}
\includegraphics[width=\linewidth]{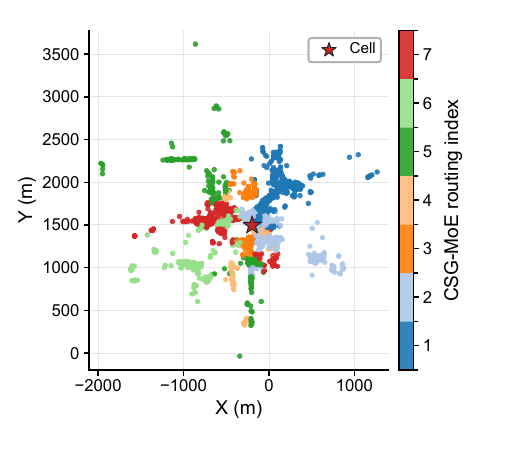}
\end{subfigure}

\caption{\textbf{Spatial organization of expert routing for a representative cell.} The same MR locations are shown in all panels, and the red star marks the serving cell. \textbf{a,} Observed RSRP distribution. \textbf{b,} Strongest-beam assignments, revealing directionally organized propagation regions around the cell. \textbf{c,} Expert assignments produced by vanilla MoE without the auxiliary strongest-beam prediction objective. The routing is concentrated in a small subset of experts and exhibits substantial spatial mixing. \textbf{d,} Expert assignments produced by CSG-MoE. The assignments form more contiguous directional regions that correspond more closely to the strongest-beam organization. Expert indices denote categorical routing identities and do not imply an ordinal relationship. Together with the reduction in reconstruction MAE from 4.98\,dB to 4.72\,dB after introducing CSG-MoE (Extended Data Table~\ref{tab:ed3_ablation}), this qualitative comparison is consistent with the auxiliary objective promoting more physically organized expert routing.}
\label{fig:ed_csg_moe}
\end{figure*}

% \section*{Acknowledgements}

% \section*{Author contributions}

% \section*{Competing interests}

\end{document}